\documentclass[preprint]{aastex}
\usepackage{emulateapj5}
\input psfig.tex
\def\etal{\emph{et al.}\ }

\def\kms{km s$^{-1}$}
\def\apj{ApJ}
\def\aj{AJ}
\def\mnras{MNRAS}
\def\apjs{ApJS}

\def\aa{{A\&A}}
\def\aas{{ A\&AS}}
\def\aj{{AJ}}
\def\al{$\alpha$}
\def\bet{$\beta$}
\def\amin{$^\prime$}

\def\apj{{ApJ}}
\def\apjs{{ApJS}}
\def\asec{$^{\prime\prime}$}

\def\deg{$^{\circ}$}

\def\e#1{$\times$10$^{#1}$}
\def\etal{{et al. }}

\def\flux{erg s$^{-1}$ cm$^{-2}$}

\def\lamb{$\lambda$}
\def\lum{erg s$^{-1}$}
\def\micron{{$\mu$m}}
\def\mnras{{MNRAS}}
\def\nat{{Nature}}
\def\pasp{{PASP}}

\def\percm2{cm$^{-2}$}

\def\lax    {${_<\atop^{\sim}}$ }
\def\gax    {${_>\atop^{\sim}}$ }

\def\hii{\ion{H}{2}}
\def\oiii{[\ion{O}{3}]}

\def\oi{[\ion{O}{1}]}
\def\nii{[\ion{N}{2}]}

\def\sii{[\ion{S}{2}]}

\slugcomment{To Appear in {\it The Astrophysical Journal Supplements}}
\shortauthors{HO \& ULVESTAD}
\shorttitle{Radio Survey of Seyferts}

\begin{document}

\title{Radio Continuum Survey of an Optically-Selected Sample of Nearby
Seyfert Galaxies}

\author{Luis C. Ho}
\affil{The Observatories of the Carnegie Institution of Washington, 813 Santa 
Barbara St., Pasadena, CA 91101-1292; lho@ociw.edu}

\and 

\author{James S. Ulvestad}
\affil{National Radio Astronomy Observatory, P. O. Box 0, 1003 
Lopezville Road, Socorro, NM 87801; julvesta@nrao.edu}

\begin{abstract}
We have used the Very Large Array (VLA) to conduct a survey for radio 
continuum emission in the sample of 52 Seyfert nuclei selected from the
optical spectroscopic galaxy catalog of Ho, Filippenko, and Sargent.  This 
Seyfert sample is the most complete and least biased available, and, as such, 
it will be useful for a variety of statistical analyses.  Here we present the 
observations, measurements, and an atlas of radio maps.

The observations were made at 6~cm in the B-array and at 20~cm in the 
A-array, yielding matched angular resolutions of $\sim$1\asec.  We detected 
44 objects (85\%) at 6~cm and 37 objects (71\%) at 20~cm above a 3 $\sigma$ 
threshold of 0.12~mJy beam$^{-1}$.  The sources have a wide range of radio
powers ($P\,\approx$ 10$^{18}$--10$^{25}$ W Hz$^{-1}$), spectral indices 
($\alpha_6^{20}\,\approx$ $+$0.5 to --1), and linear sizes ($L\,\approx$ 
few tens pc -- 15 kpc).  The morphology of the radio emission is 
predominantly that of a compact core, either unresolved or slightly resolved, 
occasionally accompanied by elongated, jet-like features.  Linearly-polarized 
emission was detected at 6~cm in 12 sources, 9 of which were also 
detected at 20~cm.

\end{abstract}

\keywords{galaxies: active --- galaxies: jets --- galaxies: nuclei --- 
galaxies: Seyfert --- galaxies: structure --- radio continuum: galaxies}

\section{Introduction}
Radio emission is one of the most distinctive attributes of the active 
galactic nucleus (AGN) phenomenon.  It varies widely in intrinsic strength and 
in outward appearance, ranging from the conspicuous large-scale jets in 
powerful radio galaxies (Fanaroff \& Riley 1974), to the less energetic radio 
outflows seen in Seyfert galaxies (e.g., Ulvestad \& Wilson 1989), to the 
more mundane compact, radio cores commonly found in elliptical (Sadler, 
Jenkins, \& Kotanyi 1989; Wrobel \& Heeschen 1991; Slee et al. 1994) and 
spiral nuclei (van der Hulst, Crane, \& Keel 1981; Sadler et al. 1995).

Aside from its ubiquity among AGNs, radio continuum emission effectively 
traces nuclear activity because it is impervious to effects which complicate
observations at shorter wavebands, namely dust obscuration in the ultraviolet, 
optical, and even near-infrared (IR) wavelengths, and photoelectric absorption 
at X-ray energies.  Modern radio interferometers, moreover, can routinely and 
efficiently deliver sensitive, high-angular resolution maps, making 
them particularly attractive instruments to survey large samples of objects.  
During the past 25 years, radio continuum surveys have contributed fruitfully
to our understanding of AGNs, especially of more nearby objects such as 
Seyfert galaxies.  Among other things, these studies have elucidated the 
pervasiveness of jet-like outflows in AGNs with a wide span of intrinsic 
power; they have delineated the gross morphology and fine structures of 
nuclear regions on scales from $\sim$1--10$^3$ pc; they have highlighted 
the intricate interplay between the radio-emitting plasma and the dynamics and
energetics of the optical emission-line regions; and they have exploited the 
directional information furnished by the radio jets to make inferences on the 
orientation of the accretion disk.  Discussions of some of these developments 
have been given by Wilson (1997), Bicknell et al. (1998), and Kinney et al. 
(2000), among others.

Many AGN investigations are statistical in nature.  Some issues raised by 
previous radio studies of Seyfert galaxies include: (1) determination of 
their radio luminosity functions (Meurs \& Wilson 1984; Edelson 1987; Ulvestad 
\& Wilson 1989; Rush, Malkan, \& Edelson 1996); (2) comparison between the 
radio properties of subtypes of Seyferts in the context of predictions from 
evolutionary or unification models (de Bruyn \& Wilson 1978; Meurs \& Wilson 
1984; Edelson 1987; Ulvestad \& Wilson 1984a, 1984b, 1989; Ulvestad 1986; 
Giuricin et al.  1990; Ulvestad, Antonucci, \& Goodrich 1995; Kukula et al. 
1995; Morganti et al. 1999; Nagar et al. 1999); and (3) comparison between 
the radio properties of Seyferts and other classes of emission-line nuclei 
(Heckman et al. 1983; Keel 1984; Giuricin, Mardirossian, \& Mezzetti 1988; 
Sadler et al. 1995).  The fidelity of the conclusions of such studies often
is limited not by sample size --- by now substantial numbers of Seyferts 
have been imaged at radio wavelengths --- but by more subtle selection 
effects inherent in most, perhaps all, samples of Seyferts galaxies.  

In an effort to circumvent some of these difficulties, we have undertaken 
a new radio continuum survey of an optically selected sample of 52
nearby Seyfert galaxies using the Very Large Array (VLA){\footnote{The VLA is 
operated by the National Radio Astronomy Observatory, a facility of the 
National Science Foundation operated under cooperative agreement by Associated 
Universities, Inc.}}.  The merits of our new sample, derived from a 
recently completed, extensive optical spectroscopic survey of nearby galaxies, 
are described in Section 2; a comparison of our sample with other major 
extant surveys is given in Appendix A.  Details of the observations, 
calibrations, and data reductions can be found in Section 3.  Section 4 
presents maps of all the galaxies and summarizes the principal source 
parameters in tabular form, including optical data pertinent to subsequent 
analysis, which will be given in a separate paper by Ulvestad \& Ho (2001).  A 
description of the radio properties of each individual object is deferred to 
Appendix B.  Some general statistical results of the survey are noted in 
Section 5.  

\section{The Palomar Seyfert Sample}

The sample of Seyfert galaxies in this paper is derived from the Palomar 
optical spectroscopic survey of nearby galaxies (Ho, Filippenko, \& Sargent 
1995).  In brief, the Palomar 200-inch telescope was employed to take 
moderate-dispersion, high-quality spectra of 486 bright ($B_T\,\leq$ 12.5 
mag), northern ($\delta\,>$ 0\deg) galaxies selected from the Revised 
Shapley-Ames Catalog of Bright Galaxies (Sandage \& Tammann 1981), with the 
primary aim of conducting an accurate census of the AGN population in the 
nearby ($z\,\approx 0$) universe.  The Palomar survey, the most sensitive 
of its kind (see Ho 1996 for a comparison with previous optical studies), 
produced a comprehensive, homogeneous catalog of spectral classifications of 
nearby galaxies (Ho et al. 1997a, 1997c).  The median distance\footnote{In 
this and subsequent papers, we adopt the distance estimates of Tully (1988), 
who assumes a Hubble constant of $H_{\rm 0}$ = 75 \kms\ Mpc$^{-1}$.} of 
the sample galaxies is 20.4 Mpc, with an interquartile range of 10.4
Mpc.  The completeness of the Palomar survey follows that of the Revised
Shapley-Ames Catalog, which, for high-surface brightness galaxies, is complete 
to $B_T$ = 12.0 mag and 80\% complete to $B_T$ = 12.5 mag (Sandage, Tammann, 
\& Yahil 1979).  A significant fraction of the survey galaxies have 
been classified as AGNs or AGN candidates (Ho et al. 1997b) ---  20\% LINERs, 
13\% LINER/\hii\ ``transition objects,'' and 11\% Seyferts.  The 52 Seyferts 
(Table~1), the subject of this paper, contain many familiar, bright objects, 
but there are also many which are not well known, which are not usually 
regarded as ``active,'' or whose classification has been revised.  

The spectroscopic classification system is described at length in Ho et al. 
(1997a).  Emission-line nuclei fall into one of two categories, \hii\ nuclei 
(powered by stars) and AGNs (powered by black-hole accretion), according to 
the relative strength of the low-ionization optical forbidden lines (\oi\ 
\lamb\lamb 6300, 6364, \nii\ \lamb\lamb 6548, 6583, \sii\ \lamb\lamb 6716, 
6731) compared to the hydrogen Balmer lines.  The ionization state of the 
narrow-line gas in AGNs, as measured through the ratio \oiii\ \lamb 
5007/H\bet, distinguishes LINERs (\oiii/H\bet\ $<$ 3) from Seyferts 
(\oiii/H\bet\ $\geq$ 3).  Finally, the presence or absence of kinematically 
distinct broad (linewidths ranging from one to several thousand \kms) 
permitted lines separates AGNs into the ``type~1'' and ``type~2'' varieties, 
respectively, while the relative strength of the broad component of the 
hydrogen Balmer lines further splits the type~1 objects into finer 
subdivisions (type 1.0, 1.2, 1.5, 1.8 and 1.9; see Osterbrock 1981).  

Table~1 (col. 2) lists the classifications of the 52 Seyferts in our sample.  
It includes, in addition, two objects (NGC~4203 and NGC~4450) classified as 
LINER~1.9 sources by Ho et al. (1997a) which were observed as part of our 
program for different reasons.  We present their data here 
for completeness, but we will exclude them from further analysis.  Three 
additional objects (NGC~1068, 1358, and 1667) do not meet the formal definition
of the Palomar sample because of their southern declination.  These galaxies 
are included in the complete statistical analysis.  The 
symbols have the following meaning: L = LINER, T = ``transition object'' 
(LINER + \hii\ nucleus), and S =  Seyfert.  Classifications deemed to be 
uncertain or highly uncertain are followed by a single and double colon, 
respectively.  In a few cases the classification assignment is ambiguous, 
and more than one is given, with the adopted classification listed first.  
NGC~4579, for example, falls near the somewhat arbitrary boundary between 
a Seyfert and a LINER, but because it is officially on the Seyfert side,
it is classified as ``S1.9/L1.9''; the ``1.9'' designation means that broad 
H\al, but not broad H\bet, has been detected.  There are 30 type~2 and 22 
type~1 objects.  

The rest of the entries in Table~1 are as follows: col. (3) Hubble type; 
col. (4) morphological type index ($T$); col. (5) total apparent $B$ magnitude
of the galaxy ($B_T$); col. (6) adopted distance ($D$); cols. (7) and (8)
optical position in epoch J2000.  Data for cols. (2)--(6) are taken from the
compilation of Ho et al. (1997a).  The optical positions with sub-arcsecond
accuracy come from Clements (1981, 1983) and Argyle \& Eldridge (1990), and
the rest are taken from Cotton, Condon, \& Arbizzani (1999).

Because of the proximity of the objects (median $D$ = 20.4 Mpc; Fig.~1{\it a\/})
and the sensitivity of the Palomar survey to weak emission lines, it is not 
surprising that our sample contains a significant number of low-luminosity 
sources.  Figure~1{\it d\/} shows the distribution of extinction-corrected 
luminosities for the narrow component of the H\al\ line; with the exception 
of the possibly anomalous object NGC~185 (see Appendix~B), $L$(H\al) ranges 
from $\sim$1\e{38} to 3\e{41} \lum, with a median value of 5\e{39} \lum.  By 
contrast, Markarian Seyferts have typical $L$(H\al) $\approx$ 
$10^{40}-10^{42}$ \lum\ (Dahari \& De Robertis 1988).  Among ``classical'' 
Seyferts, NGC~4051 holds the record for having the lowest optical continuum 
luminosity ($M_B$ = --16 
mag, adjusted to $H_{\rm 0}$ = 75 \kms\ Mpc$^{-1}$; V\'eron 1979), and in 
previous surveys, the optical luminosity function of Seyferts does not extend 
fainter than $M_B\,\approx$ --18 mag (e.g., Huchra \& Burg 1992).  Although 
nuclear continuum magnitudes are not yet available for all the objects in our 
sample, a few examples are illustrative: NGC~3031 ($M_B$ = --11.6 mag; Ho, 
Filippenko, \& Sargent 1996); NGC~4395 ($M_B$ = --9.8 mag; Filippenko, Ho, \& 
Sargent 1993); NGC~4579 ($M_B$ = --11.8 mag; Barth et al. 1996); NGC~4639 
($M_B$ = --12.5 mag; Ho et al. 1999a).  The host galaxies, on the other hand, 
are generally luminous (median $M_{B_T}^0$ = --20.7 mag; Fig.~1{\it b\/}), 
bulge-dominated (median $T$ = 2.0, or Hubble type Sab; Fig.~1{\it c\/}) galaxies.

The Palomar sample of Seyferts has several merits compared to other samples
(see Appendix A).  First, as an optically selected sample with complete
spectroscopic identification, it is less susceptible to strong biases as in 
the samples selected in the ultraviolet or IR.  (See \S\ A.2 for a 
discussion of selection biases in Seyfert samples.)  Unlike other 
optically selected samples, such as that extracted
from the CfA redshift survey (Huchra \& Burg 1992) or from heterogeneous
literature compilations (Maiolino \& Rieke 1995), the Palomar Seyferts are
derived from a survey with uniform, high-quality data, classified with
well defined and objective criteria.  Particular care was taken to
account for the severe contamination of the nuclear signal by the starlight of
the host galaxy.  Second, the close distances of the
objects offer a number of advantages.  In addition to maximizing the
achievable linear resolution and signal-to-noise ratio (S/N), the relative
proximity of the galaxies enhances the contrast between the nucleus and the
bright, surrounding bulge of the host galaxy.  This is a crucial factor
in detecting weak nuclei. Third, our sample covers a large range of AGN
luminosities, making it ideal for exploring possible trends with AGN power.
Fourth, the parent Palomar survey is sufficiently large that it contains all
nuclear spectral classes and accurate statistical representation of each
class, and so comparative analyses among different subsamples can be done
in a straightforward manner.  Finally, the Palomar objects are currently the
focus of several other studies, thus ensuring the availability of a rich
database for multiwavelength investigations.

\section{Observations and Data Reductions}

We observed the galaxies during two sessions using the VLA (see Table~2 
for a summary).  On 1999 August 29, while the array was in its A-configuration 
(Thompson et al. 1980), observations were acquired at a central sky frequency
of 1.425~GHz (L~band; 20~cm).  Matched-array data were taken with the 
B-configuration at 4.860~GHz (C~band; 6~cm) on 1999 October 31.  The weather 
conditions were excellent during both observing sessions.  To achieve maximum 
sensitivity, we used both intermediate frequencies, each of width 50~MHz, 
separated by 50~MHz.  Since we are mainly concerned with the emission from 
the nucleus, which was placed at the phase-tracking center, bandwidth 
smearing (Bridle \& Schwab 1999) is not a problem despite the somewhat 
large bandwidth.  The excellent sensitivity of the current 6 and 20~cm 
receivers of the VLA routinely permits high S/N maps 
to be made in relatively short integration times.  Each target was 
observed for 15--18 min, immediately preceded and followed by a short 
($\sim$3 min) exposure of a nearby phase calibrator to calibrate the antenna 
gains and phases.  More than half the phase calibrators used have position
accuracies of 0\farcs01 or better, either from VLBI observations (Ma et al. 
1998; Eubanks et al. 1998) or from VLA observations (Patnaik et al. 1992;
Browne et al. 1998); the rest of the calibrators have position accuracies of 
$\sim$0\farcs1 (Wilkinson et al. 1998; VLA staff, unpublished).  The 
self-calibration typically moved source positions by less than 0\farcs01, 
indicating that atmospheric phase irregularities had a relatively small effect 
on source positions.  Therefore, in most cases, the measured positions are 
accurate to 0\farcs1 or better. For a few very weak sources, the position can 
only be determined with an accuracy of approximately the beam size divided
by the S/N of the detection, which can be as large as 0\farcs3 for a 
5-$\sigma$ detection.  We tied the absolute fluxes to the VLA scale by 
observing 3C~286, whose flux densities at 6 and 20~cm were assumed to be 7.49 
and 14.75~Jy\footnote{1~Jy = 10$^{-23}$ \flux\ Hz$^{-1}$.}, respectively.  
Errors in the quoted flux densities are dominated by the uncertainty in 
setting the absolute flux scale, conservatively estimated to be 5\%.

Antenna polarization leakage was calibrated by observing the calibrator 
1313$+$675 through a wide range of parallactic angles.  The complex 
instrumental polarization of the antennas, determined to an accuracy of 
$\sim$0.2\%, was removed from the cross-polarized visibility data, and the 
right and left circularly-polarized channels were aligned by assuming that the 
position angle (P.A.) of the electric vector of 3C 286 was $+$33\deg\ at 
both 6 and 20~cm.

Two of the galaxies in our sample (NGC~3031 and NGC~7743) were observed 
with wrong coordinates and were missed.  Fortunately, both sources have been
observed previously with the VLA in setups similar to that of our program, and
we were able to retrieve their data from the archives and reprocess them in 
a way consistent with our own observations.  The NGC~3031 data were originally 
published by Kaufman et al.  (1996), while the NGC~7743 data were published by
Wrobel \& Heeschen (1991) and Nagar et al. (1999).  For NGC~7743, the 6~cm 
data set comes from the C-configuration, since the only existing 
B-configuration data also were obtained with an incorrect position.

After editing and calibration, the visibilities were Fourier transformed into 
maps of sky brightness using the algorithm IMAGR in AIPS (van Moorsel, 
Kemball, \& Greisen 1996).  The sidelobes of the synthesized beam were removed 
from the ``dirty'' maps by applying the deconvolution ``CLEAN'' procedure of 
H\"ogbom (1974), as modified by Clark (1980), to ``clean boxes'' placed around 
sources stronger than $\sim$0.5~mJy within the primary-beam area.  The number 
of iterations was set so that the minimum ``clean component'' reached 0.12~mJy 
beam$^{-1}$, $\sim$3 times the theoretical thermal noise limit of the maps.   
We constructed images at different resolutions by applying appropriate tapering 
functions to the visibilities.  Three sets of total-intensity (Stokes $I$) 
maps were made, at full resolution and with Gaussian tapering functions 
falling to 30\% at 80 and 50 k$\lambda$; these correspond to synthesized
beams of $\Delta \theta\,\approx$ 1\farcs1, 2\farcs5, and 3\farcs6, 
respectively.  The untapered maps were ``uniformly weighted'' to yield 
the highest angular resolution, at the expense of a slight degradation in 
sensitivity.  The tapered maps, ``naturally weighted'' for maximum 
sensitivity, have dimensions comparable to the half-power beam width (HPBW) of 
the antenna primary beam (8\farcm 4 at 6~cm, 28\farcm 1 at 20~cm).  They were 
used to assess the possible presence of extended emission and 
to evaluate whether sources far from the map center might be introducing 
sidelobe contamination near the nucleus.  In our chosen configurations and 
frequencies, the shortest $u$-$v$ spacings are $\sim$3 k$\lambda$, and so our
images are sensitive only to structures with angular scale \lax 1\amin.  

The theoretical root-mean square (rms) noise of our snapshot images is 
expected to be $\sim$0.04~mJy beam$^{-1}$, a level attained in most of the maps 
where sidelobe contamination, either from the nucleus itself or from nearby 
background sources, was not severe.  For sources with sufficiently high S/N, 
the dynamic range of interferometer maps can be further improved through 
self-calibration to remove antenna-based phase and amplitude errors (Cornwell 
\& Fomalont 1999).  Using the initially deconvolved maps as input models to 
the CALIB algorithm, the S/N of a number of objects improved dramatically after 
application of several cycles of self-calibration.  In a few cases a strong 
background source within the primary-beam area was used as the model.  For 
NGC~1167 and NGC~1275, the radio emission was strong enough that the dynamic 
range was still limited, probably by baseline-based errors.  Correcting such 
errors by self-calibration significantly improved the NGC~1167 images, but had 
little effect on NGC~1275.

In addition to the total-intensity images, we also made full-resolution,  
naturally-weighted maps of the Stokes $Q$ and $U$ intensities, combinations of 
which resulted in maps of the amplitude, 
$I_{\rm pol}\,\equiv\,\sqrt{Q^2 + U^2}$, and the electric vector 
P.A., $\chi\,\equiv\,1/2\, {\rm tan}^{-1}(U/Q)$ of the linearly-polarized 
intensity.  The typical rms noise of the $I_{\rm pol}$ maps is $\sim$0.025~mJy 
beam$^{-1}$.  The 3 $\sigma$ detection threshold, after correcting for the 
known positive bias of $I_{\rm pol}$ (e.g., Wardle \& Kronberg 1974), is 
$\sim$0.079~mJy beam$^{-1}$.  However, owing to linear approximations in the 
treatment of polarization impurities in the feeds, there is also a lower limit 
of $\sim$0.2\% in detectable polarization fraction.  This limit is higher than 
the nominal detection threshold for components with peak intensities greater 
than $\sim$40 mJy~beam$^{-1}$.

\section{Maps and Source Parameters}

We present the survey data as a series of greyscale images superposed with 
contours.  The two representations emphasize different details.  The maps for 
all 54 objects, including eight nondetections, are shown in Figures 2--15.  
Each figure (except Fig.~15) displays four galaxies, arranged in the order of 
increasing NGC number.  For each galaxy, the left panel ({\it a\/}) shows the 
full-resolution, uniformly-weighted 20~cm map, the middle panel ({\it b\/}) 
shows the full-resolution, uniformly-weighted 6~cm map, and the right panel 
({\it c\/}) shows the naturally-weighted, tapered 6~cm map.  Panels ({\it a\/}) 
and ({\it b\/}) are registered, both with identical dimensions, which in nearly 
all cases are 30\asec $\times$ 30\asec.  The resolution of the tapered maps is 
either $\Delta \theta\,\approx$ 2\farcs5 or 3\farcs6, depending on which better 
illustrates the extended emission, and the dimensions of some plots have been 
optimized to display particular features in each galaxy.

The restoring beam (identical for the two full-resolution panels) is depicted
as a hatched ellipse on the lower left-hand corner of each map.   The contour 
levels of the maps are rms $\times$ (--6, --3, 3, 6, 12, 24, 48, ...), where 
the rms values for the maps are given in Table~3. The optical position of the 
galaxy is marked with a cross, the semi-major length of which corresponds to 
the uncertainties given in Table~1.  Note that the cross symbol is not visible 
on the scale of these maps for objects whose optical positions have 
sub-arcsecond accuracy.  

We have adopted a consistent method in extracting source parameters. 
The rms noise of each map is determined from a source-free, rectangular 
region.  The galaxy is considered detected if a source with a peak flux 
density $S^P\,\geq$ 3 $\times$ rms is found within the error box of 
the optical position given in Table~1.  For undetected sources, the upper 
limit is set to $S^P\,<$ 3 $\times$ rms.  Whenever possible we determined 
the source parameters (position, peak flux density, integrated flux 
density, deconvolved source dimensions, and P.A.) by fitting a 
two-dimensional Gaussian model using the task JMFIT.  The P.A. is 
measured from North through East, from 0\deg\ to 180\deg.  This procedure 
works well for sources with relatively simple, symmetric structure, such as 
most galaxy cores.  For components with more complex morphologies, or for 
weak, marginal detections, we simply integrated the signal within 
interactively defined boundaries (using the task IMEAN for rectangular boxes 
and TVSTAT for irregularly shaped regions).  In these instances the highest 
pixel value defined the peak flux density and its position, and the source 
size and P.A. were estimated manually on the image display.  A source is 
considered resolved if its deconvolved size is larger than one-half the 
beam size at full width at half maximum (FWHM) in at least one of the 
two dimensions.  Upper limits to the sizes of unresolved sources are set to 
one-half the beam size at FWHM.

Two sets of measurements are given for each galaxy, one based on the 
high-resolution, uniformly-weighted map, the other on the tapered map.
In addition to a central core, some objects show obvious additional 
complex structure such as a circumnuclear ring (e.g., NGC~1068 and 
NGC~6951), jet-like extensions (e.g., NGC~1358, 3031, 3516, 4388, 5033), and 
other amorphous protrusions (e.g., NGC~3079).  If a meaningful 
decomposition of the components could be made, they were measured separately
on the high-resolution map.  

Linear polarization was detected at 6~cm for 12 objects, of which 9 were also 
detected at 20~cm (see \S\ 5).  These objects are shown in greater detail in 
Figure~16.  The contours show maps of total intensity, as in Figures 2--15, 
and superposed are vectors representing linearly-polarized emission.

The measured quantities are collected in Table~3.  Multiple entries for the 
high-resolution map denote different components; if warranted, we generally 
give separate measurements for the entire source as well as for the central 
``core'' alone. The columns are as follows: col. (1) galaxy name; col. (2) 
figure number for maps; col. (3) FWHM of the elliptical Gaussian restoring 
beam; col. (4) P.A.  of the restoring beam; cols. (5) and (6) rms noise of the 
6~cm and 20~cm maps; cols. (7) and (8) radio position in epoch J2000; 
col. (9) peak flux density at 6~cm ($S_{6}^P$); col. (10) integrated flux 
density at 6~cm ($S_{6}^I$); col.  (11) peak flux density at 20~cm 
($S_{20}^P$); col. (12) integrated flux density at 20~cm ($S_{20}^I$); col. 
(13) spectral index between 6 and 20~cm ($\alpha^{20}_6$) defined such that 
$S^I_{\nu}\,\propto\, \nu^{+\alpha}$, where $S^I_{\nu}$ is the integrated flux 
density at frequency $\nu$; col. (14) deconvolved FWHM dimensions (major 
$\times$ minor axis, $\theta_M \times \theta_m$) of the fitted source, 
determined from an elliptical Gaussian fit; col. (15) P.A. of the fitted 
source; col. (16) peak linearly-polarized flux density at 6~cm 
($S_{\rm pol,6}^P$); and col. (17) peak linearly-polarized flux density at 
20~cm ($S_{\rm pol,20}^P$).  Additional information on individual galaxies, 
including comparisons with published data, is given in Appendix B.

We have not assigned formal error bars to the radio measurements, but here
we give a brief assessment of their magnitude.  

\begin{enumerate}

\item{The radio positions for the fitted components have a nominal accuracy 
of 0\farcs1, which is dominated by the astrometric accuracy of the phase 
calibrators.  Positions based on identification of the highest peak are 
expected to have lower accuracy, on the order $\sim$0\farcs1--0\farcs5 
(see \S\ 3).  Our judgement of the accuracy of the positions is reflected 
in the number of significant digits shown in columns 7 and 8 in Table~3.}

\item{Although formal errors for the source structural parameters ($\theta_M$, 
$\theta_m$, and P.A.) can be evaluated for the elliptical Gaussian fits (e.g., 
Condon 1997), the true errors are likely to be mainly systematic in nature, 
depending on whether the source can be well represented by the model.}

\item{The uncertainty on the flux densities can be taken to be
$\sigma_S\,\approx\,\sqrt{N{\rm rms}^2\,+\,(0.05 S)^2}$, the quadrature sum of
the rms noise covered by a source occupying $N$ beam areas and a nominal 5\%
uncertainty on the absolute flux-density scale (e.g., Weiler et al. 1986).
An unresolved source with a flux density of 1~mJy typically has 
$\sigma_S\,\approx$ 0.06~mJy.}

\item{The fidelity of spectral-index measurements depends critically on the 
geometry of the source and the $u$-$v$ coverage at the two frequencies.  We 
have attempted to minimize systematic uncertainties by observing our sample 
using scaled arrays.  Our spectral-index calculations are based on maps made 
using identical mapping parameters, and we have restored the cleaned 20~cm 
maps using the synthesized beam determined from the 6~cm maps.  Nonetheless, 
our snapshot observations are not expected to yield exactly the same $u$-$v$ 
coverages at 6 and 20~cm, especially for objects at very northerly declinations 
whose beam is highly noncircular.  (Of course, the impact of differing $u$-$v$ 
coverages is not significant for sources that are clearly unresolved.)  
Furthermore, three antennas were unavailable during the 20~cm observations, but 
all 27 were present at 6~cm.  We therefore list spectral indices only for 
source components measured from the high-resolution (untapered) maps.  The 
potentially dominant source of error, however, is that due to the 
nonsimultaneity of the observations at the two bands.  Some of the sources are 
known to vary, on a variety of timescales.  For nonvariable sources or 
near-simultaneous observations, the uncertainty on the spectral index formally 
follows from $\sigma_{\alpha}\,=\, \sqrt{(\sigma_{6}/S_{6})^2\,+\, 
(\sigma_{20}/S_{20})^2)]}/[{\rm ln}\,(4.8601\, {\rm GHz}/1.4250\,{\rm GHz})]$.
For a typical 1--5~mJy source in our survey, 
$\sigma_{\alpha}\,\approx$ 0.06--0.08.}

%\item{Finally, two sources of error enter into the uncertainty on the 
%percent polarization --- that due to calibration of the antenna polarizations 
%($\sim$0.2\%), and that due to noise in the polarization maps.  The latter 
%contribution is $\sim$0.026~mJy beam$^{-1}$, after accounting for the known 
%positive bias of $P$ (\S\ 3); for a source with a total intensity 100~mJy 
%(a typical strength of our sources with detected polarization), the 
%error in $\Pi$ amounts to only $\sim$0.026\%.  The total uncertainty on 
%$\Pi$, taken to be the quadrature sum of the two contributions, is then 
%$\sim$0.6\% for 5-mJy objects, ranging down to $\sim$0.2\% for all 
%sources stronger than 20~mJy.}

\end{enumerate}

The fields of many of the objects, especially at 20~cm, contain additional 
detectable sources located outside of the error box of the optical position of 
the nucleus.  The majority of these must be background sources, although we 
cannot rule out that a few may be associated with the galaxies themselves.  
For the sake of completeness, we provide in Table~4 a list of positions and 
integrated flux densities for all the non-nuclear sources detected in the 
uniformly-weighted maps.  Note that we have identified more background sources 
at 20~cm than at 6~cm because of the larger primary beam at 20~cm and the steep 
spectrum of most field sources.  We have not attempted to track down the 6~cm 
counterparts of all the 20~cm sources.  The positions are given only to an 
accuracy of $\sim$1\asec\ because the shapes of some of the sources are 
severely distorted (elongated) due to bandwidth smearing (see, e.g., Bridle \& 
Schwab 1999).  Whenever possible, we measure the positions from the 6~cm 
maps because bandwidth smearing is less severe at the higher frequency.  The 
flux densities have not been corrected for primary-beam attenuation.

A compilation of derived radio parameters is given in Table~5, along with 
several optical parameters which will be used in the analysis by Ulvestad \& 
Ho (2001).  The two galaxies not in the Seyfert sample (NGC~4203 and NGC~4450) 
are not included in this table.  The radio quantities listed are as follows: 
cols. (2) and (3) logarithm of the peak core ($P_{6}^{\rm core}$) and 
integrated total ($P_{6}^{\rm tot}$) power at 6~cm; cols. (4) and (5) 
logarithm of the peak core ($P_{20}^{\rm core}$) and integrated total 
($P_{20}^{\rm tot}$) power at 20~cm; col. (6) largest angular size
($\Theta$); col. (7) largest linear size ($L$); col. (8) P.A. of the source major
axis (P.A.$_{\rm rad}$); and col. (9) radio morphology class.  The
quantities listed pertain only to the ``AGN'' component of the maps.  The 
circumnuclear rings of NGC~1068 and NGC~6951, for example, have not been 
included in the measurements.  The total integrated powers were measured from 
the tapered maps, if these exhibit additional emission pertinent to the 
AGN component which has been resolved out in the untapered maps.
When determining the extent and orientation of 
the radio source, we examined both the tapered and untapered maps, 
at 6 and at 20~cm, and we adopted one final representation to make the 
measurements. The source size occasionally differed slightly among the maps, 
but in general there is good consistency.  We adopt the definitions of 
Ulvestad \& Wilson (1984a) for the radio morphology classes: ``U'' (single, 
unresolved), ``S'' (single, slightly resolved), ``D'' (diffuse), ``L'' (linear 
structure or multiply aligned components), and ``A'' (ambiguous).  
A slightly resolved source is one whose deconvolved size is $\geq$1/2 the 
synthesized-beam width.   When a morphological classification includes 
two characters, the first refers to the core and the second to the extended 
emission.  For instance, ``U+L'' indicates an unresolved (U) core within a 
linear (L) extended source.  The following optical data were taken from the 
catalog of Ho et al. (1997a): col. (10) total absolute $B$-band magnitude of 
the galaxy, corrected for internal extinction ($M_{B_T}^0$) as prescribed by
de~Vaucouleurs et al. (1991); col. (11) P.A. of the optical major axis of the
galaxy (P.A.$_{\rm opt}$); cols. (12) and (13) logarithm of the luminosity of
the \oiii\ \lamb5007 [$L$(\oiii)] and narrow H\al\ [$L$(H$\alpha$)] emission
lines (\lum), both corrected for Galactic and internal extinction; and
col. (14) FWHM of the \nii\ \lamb 6583 emission line [FWHM([N II])],
which is taken as representative of the velocity dispersion of the 
narrow-line gas.  We note that the emission-line luminosities of any 
individual object may have significant uncertainty, although statistical 
measures for the whole sample or for substantial portions thereof should be 
more reliable; see Ho et al. (1997a) for a detailed discussion of the 
limitations of the optical spectroscopic data.

\section{General Results and Summary}

The optical spectroscopic survey of Ho, Filippenko, and Sargent (1995, 1997a) 
has yielded a new, comprehensive catalog of 52 Seyfert galaxies, the most 
complete and least biased available.  We acquired moderately high-resolution 
($\sim$1\asec) 6 and 20~cm observations using the VLA to characterize the 
radio continuum properties of this sample.  This paper presents the 
observational material, along with a compilation of optical parameters to be 
used in subsequent analyses.

We have detected the vast majority of the objects in our sample.  Forty-four 
of the 52 Seyfert galaxies in our sample, 85\%, were detected at 6~cm.  The 
success rate at 20~cm is somewhat lower (37/52 or 71\%), most likely because 
of the higher noise in the maps due to confusion from background sources.  
Eight of the sources seen at 6~cm (NGC~3185, 3941, 4169, 4378, 4477, 4639, 
4698, and 5631) may be deemed marginal detections; their peak flux densities 
are on the order of 0.15--0.3~mJy beam$^{-1}$, or $\sim$3.5--6 times the rms 
noise.  In nearly all cases, however, the source is present both in the 
untapered {\it and\/} in the tapered maps, thus lending confidence to its 
reality.  Only two (NGC~4169 and NGC~5631) of these eight, on the other hand, 
were also detected at 20~cm.  Although this might suggest that these weak 6~cm 
sources are spurious, it is possible that they escaped detection at 20~cm 
because of the higher systematic noise in these maps.

The sources in our sample extend to significantly lower radio powers than 
in many previous surveys.  Figure~17 shows distributions of the core and total 
radio powers.   The weakest source has 10$^{18}$ W Hz$^{-1}$, the strongest 
$\sim$10$^{25}$ W Hz$^{-1}$.    The median powers for the detected 
sources are as follows: $P_{6}^{\rm core}$ = 1.7\e{20} W Hz$^{-1}$, 
$P_{6}^{\rm tot}$ = 3.5\e{20} W Hz$^{-1}$, $P_{20}^{\rm core}$ = 4.4\e{20} 
W Hz$^{-1}$, and $P_{20}^{\rm tot}$ = 9.5\e{20} W Hz$^{-1}$.

A wide range of morphologies is seen, but the most common is that of a single
compact source centered on the position of the optical nucleus.  Thirty-six
of the 44 objects (82\%) detected at 6~cm contain either an unresolved 
(23) or slightly resolved (13) central source.  Among these, 14 (or 32\% of 
all detected sources) have additional extended emission which can be 
considered ``linear,'' many reminiscent of jet-like or outflow structures.  
The eight marginally detected sources have simply been labeled as ``ambiguous."

A histogram of the largest linear sizes is shown in Figure~18{\it a}.  
They range from a few tens of parsecs for the nearest objects to $\sim$15 kpc 
for the most distant.  The mean of the distribution, computed using the 
Kaplan-Meier product-limit estimator (Feigelson \& Nelson 1985) to account 
for the upper limits, is 2.1$\pm$0.54 kpc.
 
While the extended components of the radio sources tend to have steep spectra, 
a nonnegligible fraction of the cores have flat or even inverted spectra 
(Fig.~18{\it b\/}).  If we define a flat-spectrum source as one with 
$\alpha_6^{20}\,\geq$ --0.30, 35\% (13/37) of the cores with measurable 
spectral indices qualify as such.  The median of the distribution of 
$\alpha_6^{20}$ is --0.46.

Finally, linearly-polarized emission was detected in 12 objects.  Nine
objects (NGC~1068, 1167, 2655, 3031, 4151, 4472, 5194, 5548, and 7479) 
were detected at 6 and 20~cm, and three (NGC 2639, 3079, and 3147) only 
at 6~cm.  Radio polarization has been reported previously in only a small 
handful of Seyferts (NGC~1068, Wilson \& Ulvestad 1983; NGC~3031, Kaufman et 
al. 1996; NGC~3079, Duric \& Seaquist 1988; NGC~5194, Crane \& van der Hulst 
1992).  Prior to this study, radio polarization has not been searched 
systematically in most Seyfert galaxies.  Often, radio observations have been 
short snapshots in which no calibrator has been observed over a wide enough 
range of parallactic angles to calibrate the antenna polarization properties.  
In the present sample, the detection of polarization is limited evidently by 
sensitivity, since the detected sources all tend to have the highest 
total flux densities.  Whereas the polarization detection rate is just 27\% 
(12/44) for all the objects detected in total intensity at 6~cm, it climbs to 
55\% (12/22) among those with $S^I_6\,\geq$ 10~mJy, and it reaches 89\% for 
the nine brightest sources with $S^I_6$ \gax 50~mJy.  With only a few 
exceptions, the strongest polarization signal within each source is generally 
{\it not\/} coincident with the galaxy core.  Instead, the polarization 
fraction is highest in the extended components, often associated with 
jet-like or outflow features.  Good examples can be seen in NGC~1068 
(Fig.~16{\it a\/}), NGC~3031 (Fig.~16{\it d\/}), NGC~3079 (Fig.~16{\it g\/}), 
and NGC~5548 (Fig.~16{\it k\/}).  And even in galaxies where the polarized 
emission is associated with a slightly resolved core, the polarization vectors
seem to trace smaller scale structure directed away from the nucleus 
(e.g., NGC~1167, Fig.~16{\it b\/}; NGC~4472, Fig.~16{\it j\/}).

A more complete description of the statistical properties of this survey and 
their astrophysical implications will be given by Ulvestad \& Ho (2001).

\acknowledgements
The research of L.~C.~H. is partly funded by NASA grant NAG 5-3556, and by NASA
grants GO-06837.01-95A and AR-07527.02-96A from the Space Telescope Science
Institute (operated by AURA, Inc., under NASA contract NAS5-26555).  
L.~C.~H. is grateful for travel support from the NRAO and thanks the VLA staff 
for their hospitality during an extended visit to Socorro in December 1999.
We thank Marianne Vestergaard for helpful comments on the manuscript, and 
Joan Wrobel for useful discussions of polarization upper limits.
This work made extensive use of the NASA/IPAC Extragalactic Database
(NED) which is operated by the Jet Propulsion Laboratory, California Institute
of Technology, under contract with NASA.

%APPENDIX
%\clearpage
\appendix

\section{Sample Considerations}

\subsection{Previous Surveys}
To fully appreciate the merits and limitations of the sample of Seyferts 
studied in this paper, it is useful to recapitulate the features of the main 
extant radio continuum surveys of Seyferts.  

\subsubsection{The Westerbork Surveys}  
The Westerbork Synthesis Radio Telescope (WSRT) was used to map at 20~cm a 
sample of $\sim$75 Seyfert galaxies and related emission-line objects chosen 
from the first nine lists of Markarian objects.  The rms noise of the maps is 
$\sim$1--1.3~mJy beam$^{-1}$, and the synthesized beam is $\sim$25\asec. The 
results of the survey are presented in de Bruyn \& Wilson (1976, 1978), Meurs 
\& Wilson (1981, 1984), and Wilson \& Meurs (1982).  Most of the objects are 
relatively distant, with $<D>\,\approx$ 250 Mpc, and those which were detected 
have typical powers of $P_{\rm 20~cm}\,\approx\, 10^{21} - 10^{24}$ W Hz$^{-1}$.

\subsubsection{The VLA Surveys}
Soon after the construction of the VLA, it was used to map in 
greater detail the brightest Markarian Seyferts which were detected earlier at 
Westerbork (Wilson \& Willis 1980; Ulvestad, Wilson, \& Sramek 1981; 
Ulvestad \& Wilson 1984a).  These observations were extended to include 
all known Seyfert galaxies observable from the VLA ($\delta\,\geq\, -45$\deg) 
within $cz\,=$ 3100 \kms\ (the ``nearby sample,'' total 25 objects; 
Ulvestad \& Wilson 1984b), and then additional objects within $cz$ = 
4600 \kms\ (the ``distance-limited sample,'' total 57 objects; Ulvestad \& 
Wilson 1989).  The majority of the VLA observations were done at 6 and 20~cm 
in the A-configuration, and the resulting maps had a typical synthesized beam 
of \lax 1\asec\ and an rms sensitivity of 0.1--0.2~mJy beam$^{-1}$.  The latest 
update to this series is published by Nagar et al.  (1999), who, in addition 
to a recessional velocity limit $cz\,<$ 7000 \kms, further restricted the 
sample to early-type host galaxies (mostly S0s).  The maps were made at 3.6 
and 20~cm, with resulting beam sizes of $\sim$0\farcs3 and $\sim$1\asec, and 
rms sensitivities of 0.04--0.09 and 0.08--0.15~mJy beam$^{-1}$, respectively.

\subsubsection{The Piccinotti et al. Sample}
The {\it HEA0-1}\ A-2 all-sky survey produced a complete sample of AGNs, the 
majority of which are Seyferts, selected in the hard X-ray (2--10 keV) band 
(Piccinotti et al. 1982).  The 28 sources accessible from the VLA 
have been detected at 6 and 20~cm with resolutions \lax 1\asec\ 
(Unger et al. 1987).

\subsubsection{The CfA Seyferts}
The CfA redshift survey produced an optical magnitude-limited ($m_{pg}\,<$ 
14.5 mag) sample of 48 Seyfert galaxies (Huchra \& Burg 1992).  Edelson (1987) 
obtained observations of this sample at 1.5~cm using the Owens Valley 
single-dish 40~m telescope (beam $\sim$1\farcm5) and at 6 and 20~cm using the 
VLA in its D-configuration (beam $\sim$ 15\asec\ and 45\asec, respectively).  
The angular resolution of these data, however, was too coarse to separate the 
nuclear emission from the host galaxy emission, which in many cases is not 
insignificant.

Kukula et al. (1995) remedied this problem by reobserving the CfA sample 
with the VLA in the A and C-configurations at 3.6~cm, producing 
high-sensitivity maps (rms $\approx$ 0.07--0.1~mJy beam$^{-1}$) at resolutions 
of 0\farcs3 and 2\farcs5, respectively.  The CfA Seyferts have distances of 
10--300 Mpc, and they emit $P_{\rm 20~cm}\,\approx\, 10^{20} - 10^{23}$ 
W Hz$^{-1}$.

\subsubsection{The 12-\micron\ Sample}
Spinoglio \& Malkan (1989) constructed a galaxy catalog selected at 12 
\micron\ from the {\it Infrared Astronomical Satellite (IRAS)} database, 
within which they identified a sample of AGNs based mostly 
on existing published optical spectroscopic classifications. The 12-\micron\ 
sample contains $\sim$60 Seyferts; the radio properties of 42 of these 
were studied by Rush et al. (1996) using the VLA at 6 and 20~cm in the 
D-configuration.  As in Edelson's (1987) treatment of the CfA sample, the 
coarse resolution of the maps potentially confuses the nuclear regions with 
significant emission from the host galaxy.  The 12-\micron\ objects have 
distances up to 300 Mpc, and so the coarse synthesized beams of 15\asec\ (6~cm) 
and 45\asec\ (20~cm) correspond to linear dimensions of 23 and 68 kpc, 
respectively.  Thean et al. (2000) recently have obtained VLA 3.6~cm maps of 
the ``extended'' 12-\micron\ sample (Rush, Malkan, \& Spinoglio 1993), 
at much higher angular resolution, at $\Delta \theta\, \approx$ 0\farcs3.

\subsubsection{The ``Warm'' {\it IRAS\/} Sample}
Luminous AGNs can be identified efficiently by their far-IR colors because 
they tend to have flatter IR spectra (``warmer'' colors) compared to 
normal or starburst galaxies (de Grijp et al. 1985, 1987, 1992; Osterbrock \& 
De Robertis 1985; Kailey \& Lebofsky 1988; Low et al. 1998; Keel, de Grijp, 
\& Miley 1988).  Schmitt et al. (2001) have analyzed new and archival 3.6~cm 
VLA A-array data for a subset of 74 nearby ($z\,\leq$ 0.031) Seyferts 
from the sample of de Grijp et al. (1987, 1992).  

\subsubsection{The PTI Studies}
The two-element Parkes-Tidbinbilla Interferometer (PTI) has a baseline of
275~km, which at 8.4~GHz  produces a fringe spacing of 0\farcs03.  A set of 
studies have capitalized on this capability of the PTI, using it to search for 
compact, high-brightness temperature cores in Seyfert galaxies.  Norris et al. 
(1988, 1990) and Roy et al. (1994) observed at 1.7 and 2.3~GHz a large sample 
of spiral galaxies, some of which are classified as Seyferts, selected by 
their {\it IRAS\/} far-IR flux.  Sadler et al. (1995) extended the 
frequency coverage to 8.4~GHz.  The rms sensitivities of the PTI data 
are 0.5--3~mJy beam$^{-1}$.

\subsubsection{Others}
For completeness, we mention three other related studies. 

Ulvestad (1986) made VLA maps of the 10 ``intermediate'' type (1.8 and 1.9) 
Seyferts (Osterbrock 1981; Cohen 1983) known as of early 1984.  The 
observations were done in the A and B-configurations, at 6 and 20~cm, yielding 
maps with resolutions of 0\farcs5--4\asec\ and rms sensitivities of 
$\sim$0.1~mJy beam$^{-1}$.

Osterbrock \& Pogge (1985) drew attention to the unusual kinematic and 
excitation characteristics of a subclass of objects since termed 
``narrow-lined Seyfert 1 galaxies.''  Ulvestad et al. (1995) examined the 
radio properties of the 17 such objects for which optical spectropolarimetric 
measurements were obtained by Goodrich (1989).  Their compilation includes 
data at 3.6, 6, and 20~cm, with resolutions ranging from 0\farcs2 to 1\farcs2.

Finally, Morganti et al. (1999) presented radio continuum maps for 29 Seyferts 
chosen from a total sample of 51 which satisfy $\delta\,<$ 0\deg\ and 
$cz\,<$ 3600 \kms, and which were considered by the authors to be ``well 
classified.''   The remaining 22 objects overlap heavily with studies 
such as those of Ulvestad \& Wilson (1984b, 1989).  VLA maps with resolution 
$\sim$1\asec\ were made at 6~cm for objects in the declination range 
--30\deg\ $<\,\delta\,<$ 0\deg; the more southern objects were observed at 
3.5~cm using the Australian Telescope Compact Array at a similar resolution.

\subsection{Selection Biases in Seyfert Samples}

The parent Seyfert samples on which many of the above radio surveys 
are based have inherent selection biases and completeness issues which 
need to be considered.

Because the Markarian survey selected objects by their ultraviolet excess, 
Seyfert galaxies derived from it are in principle biased toward members with 
unusually low dust extinction or exceptionally blue intrinsic spectra.  
Because Seyfert 2 nuclei display weaker featureless, blue continua than 
Seyfert 1 nuclei in their observed spectra (e.g., Koski 1978), any survey 
which selects objects by their ultraviolet or blue flux will have an 
overrepresentation of Seyfert 2 galaxies with intrinsically luminous, 
atypically blue nuclei, or, alternatively, Seyfert 2 galaxies with unusually 
high levels of ultraviolet emission arising exterior to the nucleus, such as 
in near-nuclear or circumnuclear star-forming regions.  Ultraviolet selected 
Seyferts, therefore, contain mismatched populations of type~1 and type~2 
objects.  This selection bias may account for the frequent reports that 
Seyfert 2s tend to have a higher incidence of nuclear star formation compared 
to Seyfert 1s (e.g., Colina et al. 1997; Gonz\'alez Delgado et al. 1998).

Spinoglio \& Malkan (1989) argue that the {\it IRAS\/} 12 \micron\ band 
minimizes wavelength-dependent selection effects for AGNs and that it carries 
an approximately constant fraction of the AGN bolometric luminosity.  The 
``warm'' {\it IRAS\/} samples selected at 25 and 60 \micron\ have the 
advantage of being less susceptible to biases by Seyfert type (Keel et al. 
1994; Kinney et al. 2000).  Nonetheless, selection by mid-IR and far-IR 
emission most likely singles out unusually dusty objects, as well as those 
which may be simultaneously experiencing elevated levels of star formation, 
either in and near the nucleus or further out, since the large {\it IRAS\/} 
beam admits substantial contributions from the large-scale emission of the 
host galaxy.  Kinney et al. (2000) note that the sample of Schmitt et al. 
(2001), for example, has a preponderance of highly inclined (edge-on) systems.
And, as recognized by de Grijp et al. (1985, 1987) and Keel et al. (1988), 
while selection by far-IR colors identifies AGNs very effectively, it by 
no means produces complete samples.  {\it IRAS}-based Seyfert samples 
are especially incomplete in low-luminosity objects because of severe 
confusion with the host galaxy.

Selection by hard X-rays in principle can yield relatively unbiased 
AGN samples.  All known classes of AGNs emit X-rays, and photoelectric 
absorption is neglible at energies greater than 1~keV for gas columns 
\lax 10$^{25}$ atoms cm$^{-2}$.  To date, however, the only all-sky 
hard X-ray survey is that by Piccinotti et al. (1982) based on the 
{\it HEA0-1}\ A-2 experiment.  Because of the shallow limit of 
the survey [$f$(2--10 keV) $>$ 3\e{-11} \flux], only a relatively 
limited number of intrinsically luminous sources are included.

An effective, albeit time-consuming, method to select AGNs is through 
optical spectroscopy of large numbers of galaxies chosen by well defined 
criteria, such as samples limited by optical magnitude.  This category 
includes the CfA redshift survey (Huchra \& Burg 1992) and the Palomar 
survey (Ho et al. 1995).  AGNs are identified by diagnostic intensity 
ratios and widths of emission lines.  This technique has two obvious 
limitations.  First, heavily obscured sources could be missed in the 
optical, depending on the spatial scale and distribution of the dust.  
For example, dust tori on parsec or sub-parsec scale can easily hide the 
broad-line region and central continuum source, as is well known in some 
Seyfert nuclei.  Nonetheless, this kind of obscuration in principle affects 
only the detailed classification of the AGN (as a ``type~1'' or ``type 2'' 
object); it would not cause the AGN to go unnoticed.  The narrow-line regions 
of Seyferts typically extend over the inner few hundred parsecs or more, 
comparable to the physical scales sampled by ground-based optical 
spectrographs.  The impact of dust distributed over galaxy-wide scales, such 
as in edge-on or interacting systems, is harder to gauge.  In general, however, 
it appears that sensitive spectra can detect AGN emission lines even 
in highly obscured systems, presumably because the dust distribution 
is patchy.  A significant fraction of ultraluminous IR galaxies, 
for instance, many of which are dusty and highly obscured, are recognized as 
AGNs (e.g., Veilleux, Kim, \& Sanders 1999).  Ho et al. (1997b) find that 
selection biases due to inclination effects do not appear to be severe
in the Palomar survey.  A second, potentially more serious complication arises 
from dilution of the AGN signal by emission from \hii\ regions that might be 
included in the spectrograph aperture.  Weaker AGNs can be outshined 
by brighter contaminating \hii\ regions, especially in late-type galaxies 
undergoing vigorous star formation.  This effect may account in part for the 
apparent dearth of AGNs among late-type galaxies in the Palomar survey 
(Ho et al. 1997b); it is unlikely to be significant in early-type (E--Sbc) 
galaxies, however, where the detection rate of AGNs is already very high 
($\sim$60\%; Ho et al. 1997b).  To summarize: although AGN samples selected 
by optical spectroscopy are not immune from selection effects, they 
appear not be very serious, especially for early-type (E--Sbc) galaxies.

We close with a few comments regarding some of the existing 
optically selected Seyfert samples.  The CfA Seyfert sample (Huchra \& Burg 
1992) is widely considered to be relatively free of systematic biases.  We 
draw attention to some aspects of this sample which introduce subtle 
selection biases that appear to have been generally unrecognized.  Huchra \& 
Burg identified their Seyfert spectra by visual 
inspection of plots following the criteria stated in Huchra, Wyatt, 
\& Davis (1982): objects whose permitted lines (H\al\ and H\bet) have 
a full width near zero intensity (FWZI) larger than 2500 \kms\ are called 
Seyfert 1s, whereas those with FWZI(H\al) = 600--2500 \kms\ and a ``large'' 
\oiii\ \lamb5007/H\bet\ ratio are dubbed Seyfert 2s.   They further define 
LINERs as objects with ``strong'' \oi\ \lamb6300 emission. Several criticisms 
can be raised against these selection criteria.  First, the FWZI is a 
highly subjective parameter which depends sensitively on the S/N of a
spectrum.  The original CfA redshift survey, on which the Huchra \& Burg 
catalog is based, was conducted largely with a photon-counting 
Reticon detector, a unique attribute of which is that the spectrum can be 
monitored in real time as photons accumulate.  Since the primary 
objective was to obtain galaxy redshifts, the integration time was truncated 
once spectral features suitable for redshift determination could be discerned
in the spectrum (Tonry \& Davis 1979).  The resulting nonuniform S/N of the 
data must make detection of broad, low-contrast wings on emission lines 
particularly difficult.   Second, defining the FWZI for the narrow component 
of H\al, in the case of the Seyfert 2s, is especially treacherous, since in 
AGNs H\al\ is quite heavily blended with the flanking \nii\ \lamb\lamb 6548, 
6583 lines, and the lines often have non-Gaussian, asymmetric 
profiles near the base  (see Ho et al. 1997c for examples).  Only those
Seyfert 2s having the broadest, most extended wings observed with spectra of
very high S/N would have made it into the final sample.  Third, 
it is well known that the narrow lines of Seyferts span a substantial range 
of line widths, which are positively correlated with, among other things, 
the optical and radio luminosity of the nucleus (Phillips, Charles, \& Baldwin 
1983; Whittle 1985, 1992; Ho 1996).  Selecting Seyferts based on an 
arbitrary lower threshold in narrow-line width, therefore, translates into 
selection by minimum optical and radio AGN power.  Moreover, the lower bound 
of FWZI(H\al) = 600 \kms\ for Seyfert 2s chosen by Huchra et al. (1982) 
corresponds roughly to FWHM(H\al) = 300 \kms, as judged from typical observed 
spectra.  There are many Seyferts, of type~1 and type~2, whose narrow emission 
lines have widths smaller than this value.  In the Palomar sample studied 
here, for example, the median FWHM for \nii\ \lamb6583, which serves as an 
acceptable surrogate for H\al\ (see Ho et al. 1997a), is 275 \kms, with an 
interquartile range of 85 \kms.   And finally, the CfA catalog considerably 
blurs the distinction between LINERs and Seyferts because of the somewhat 
subjective, nonstandard manner in which the classification criteria have been 
chosen.

We conclude, from the above discussion, that the CfA Seyfert sample is likely 
to be complete only for relatively bright objects with very conspicuous 
broad emission lines.  In the case of Seyfert 2s, there is a preponderance of 
intrinsically luminous and more radio-powerful objects.  A concrete comparison 
between the Palomar and CfA Seyfert samples may be illuminating.  Of the 52 
Seyferts in the Palomar sample, 39 formally fall within the limits of the CfA 
redshift survey, but only 12 (31\%) --- all classified as type~1 objects in 
the Palomar survey --- were called Seyferts.  Another nine objects which we 
classify as Seyferts were labeled LINERs. Even if we dispense with the 
arguably nebulous distinction between LINERs and Seyferts, the detection rate 
of the Palomar objects is still only 54\% (21/39).

Lastly, we note that Seyfert samples drawn from heterogeneous literature 
sources are likely to contain complicated, ill defined incompleteness 
problems.  This cautionary note would apply to studies which selected Seyferts 
based on spectral classifications ``known'' at the time.  The samples 
in this category include those of Ulvestad \& Wilson (1984b, 1989), Spinoglio 
\& Malkan (1989), Rush et al. (1993), Nagar et al. (1999), and Morganti et al. 
(1999).

\section{Notes on Individual Objects}

This Appendix provides a short description of each object, along with 
some remarks based on previously published radio observations.  Note that 
the literature references are not meant to be comprehensive; we cite only the 
most pertinent information.  The optical spectroscopic classification, taken 
from Ho et al. (1997a), is given in parentheses after the object name.  See 
Section 2 for more details on the classification notation.

\noindent {\it NGC~ 185 (S2)}. ---  Not detected.  Heckman, Balick, \& 
Crane (1980) also did not detect the galaxy at 6~cm ($S_6\,<$ 1~mJy; $\Delta 
\theta$ = 1\farcs7) or 20~cm ($S_{20}\,<$ 5~mJy; $\Delta \theta$ = 4\asec).
Although the weak emission lines in this object formally place it in the 
category of Seyferts, it is probable that this galaxy does not contain a 
genuine active nucleus.  Instead, its emission lines may be powered by stellar 
processes.  NGC~185 is one of the companions of M~31, and it is 
morphologically classified as a dwarf spheroidal galaxy, one with a diffuse 
center lacking a clear nucleus.

\noindent {\it NGC~ 676 (S2:)}. --- Not detected.  No previous radio 
observations. 

\noindent {\it NGC~ 777 (S2/L2::)}. --- The core appears to be slightly 
resolved at both frequencies along P.A. = 134\deg.  There are no previous 
high-resolution radio observations.

\noindent {\it NGC~1058 (S2)}. --- Not detected. Cowan, Henry, \& Branch 
(1988) obtained deep radio observations of the site of SN~1961V, but the 
nucleus of the galaxy, which lies within 1\farcm 3 of the supernova, was not
detected at 6~cm ($S_6\,<$ 0.09~mJy; $\Delta \theta$ = 0\farcs9) or 20~cm 
($S_{20}\,<$ 0.12~mJy; $\Delta \theta$ = 0\farcs9).

\noindent {\it NGC~1068, M 77 (S1.9)}. --- This well known source has 
been extensively studied by many authors (e.g., Wilson \& Ulvestad 1982b, 1983;
Ulvestad, Neff, \& Wilson 1987; Kukula et al. 1995; Gallimore, Baum, \& O'Dea 
1996).  In addition to the prominent lobes associated with the AGN, our maps 
show traces of the 30\asec-diameter ($\sim$2 kpc) circumnuclear ring. Our 
measurement of the 6~cm core is in good agreement with the results of van der 
Hulst et al. (1981), whose data had a similar resolution as ours.  
Wilson \& Ulvestad (1983) reported linearly-polarized 6~cm emission 
associated with the core and the northern lobe of NGC~1068.  We detected 
these components ($S^I_{\rm pol, 6}$ = 0.91~mJy and 5.1~mJy, respectively), 
and, in addition, we find significant polarization ($S^I_{\rm pol, 6}$ = 
5.3~mJy) associated with the low-surface brightness feature to the northeast of 
the core (Fig. 16{\it a}).   At 20~cm, the polarized signal near the core and 
in northern lobe can still be seen ($S^I_{\rm pol, 20}$ = 1.2~mJy and 1.1~mJy, 
respectively), although the northeastern feature appears much weaker.

\noindent {\it NGC~1167 (S2)}. --- The core appears clearly extended in 
our images.  Our flux densities are in good agreement with those of Fanti 
et al. (1986), who quote $S_6$ = 939~mJy ($\Delta \theta$ = 0\farcs4) and 
$S_{20}$ = 1780~mJy ($\Delta \theta$ = 1\farcs4).  The higher resolution 6~cm 
data of Fanti et al. resolved the extended core into two components, which are 
also seen in the 2~cm map ($\Delta \theta$ = 0\farcs4) of Bridle \& Fomalont 
(1978).  We detected linearly-polarized emission at 6 ($S^I_{\rm pol, 6}$ = 
1.8~mJy) and 20~cm ($S^I_{\rm pol, 20}$ = 2.1~mJy).  The polarization vectors, 
especially at 6~cm, appear to trace the slight extensions seen in the 
total-intensity image (Fig. 16{\it b}).  The 20~cm polarization at the 
total-intensity peak is near the detection threshold of 0.2\%, but we believe 
it to be real due to its continuity with the clearly extended polarized
emission to the northeast.

\noindent {\it NGC~1275 (S1.5)}. --- The radio source 3C~84 associated with 
NGC~1275 possesses well known jets which have been documented on a variety of 
scales (Pedlar et al. 1990; Dhawan, Kellerman, \& Romney 1998; Silver, Taylor,
\& Vermeulen 1998; Walker et al. 2000).   The only measurable polarization
was on the radio core, and well below the threshold of 0.2\%, so
there is no significant polarization detection for this galaxy.

\noindent {\it NGC~1358 (S2)}. --- Both Ulvestad \& Wilson (1989) and 
Nagar et al. (1999) have noted that the core of NGC~1358 may be 
slightly extended.  Our full-resolution 20~cm and tapered 6~cm maps show quite 
clearly that the nucleus is straddled by a jet-like linear feature with a 
total extent of $\sim$8.6 kpc along P.A. = 101\deg.  We note that the 
optical major axis of the galaxy is at P.A. = 165\deg, and so the 
extended radio emission is likely to be associated with the AGN rather 
than the galaxy disk.

\noindent {\it NGC~1667 (S2)}. --- Ulvestad \& Wilson (1989) detected a 
core at 6 and 20~cm with flux densities comparable to ours, and Thean et al. 
(2000) give a similar value of $S_6$ = 1.5~mJy ($\Delta \theta$ = 0\farcs3).  
Our tapered 6~cm map shows, in addition, a low surface brightness two-armed 
spiral feature, with a total linear extent of 14.5 kpc along P.A. $\approx$ 
0\deg.  The tapered 20~cm image only shows the central core and the 
high-surface brightness region of the southern spiral arm.

\noindent {\it NGC~2273 (S2)}. --- Our maps show the central source to 
be slightly extended in both bands, more so at 20~cm than at 6~cm.  The 
20~cm map of Ulvestad \& Wilson (1984b), made with a resolution of 1\farcs4, 
agrees well with ours.  Their higher resolution 6~cm map ($\Delta \theta$ =
0\farcs4), however, shows that the central source resolves into two 
components separated by 0\farcs9 along the east-west direction.  
The $\Delta \theta\,\approx$ 4\asec\ 6~cm WSRT map of Baum et al. (1993) 
contains an extra resolved component not visible in our tapered 6~cm map.

\noindent {\it NGC~2639 (S1.9)}. --- We detect an unresolved core with an 
inverted spectrum ($\alpha_6^{20}$ = 0.47).  The higher resolution 6~cm map of 
Ulvestad \& Wilson (1989) shows that the structure is suggestive of an 
incompletely-resolved triplet along P.A. $\approx$ 110\deg.  This is 
consistent with other observations: 6~cm VLBA ($\Delta \theta$ = 2 mas; Wilson 
et al. 1998), 6~cm VLA ($\Delta \theta$ = 0\farcs6; Condon et al. 1982), and 
3.6~cm VLA ($\Delta \theta$ = 0\farcs3; Thean et al. 2000).  Our 6~cm flux 
density (182~mJy) is significantly higher than that measured by the above-cited 
higher resolution studies (50--55~mJy).  This can be interpreted as a
resolution effect, although it can also result from source variability
(Wilson et al. 1998).  We further note that our maps show two additional 
``ears'' of emission emerging on either side of the core along the north-south 
direction, roughly perpendicular to the major axis of the central component 
resolved by Ulvestad \& Wilson (1989).  Although these features are faint 
($S_6\,\approx$ 0.8 and 0.7~mJy for the northern and southern component, 
respectively), they appear to be real because they are present in the 
full-resolution 6 and 20~cm maps, and the northern component barely peaks 
through in the tapered 6~cm map.  Because of their uncertain relation to the 
AGN, however, we do not include these features in the radio morphological 
classification.  Weak polarized emission is present at 6~cm ($S^I_{\rm pol, 6}$ 
=  0.61~mJy), but not at 20~cm ($S^P_{\rm pol, 20}\,<$ 0.086~mJy beam$^{-1}$).  
Most of the signal is associated with the core, but, quite interestingly, the 
north-south ``ears'' also appear to be polarized (Fig.~16{\it e\/}).

\noindent {\it NGC~2655 (S2)}. --- Keel \& Hummel (1988) have reviewed the 
radio properties of NGC~2655.  They noticed that apart from the central 
core, which had previously been studied by van der Hulst et al. (1981) and 
Hummel, van der Hulst, \& Dickey (1984), there exists a jet-like feature 
located $\sim$15\asec\ (1.8 kpc) to the southeast.  This patch of radio 
emission is evidently associated with an optical emission-line region, which 
could plausibly be powered by the radio plasma.  Our data reveal an additional 
component, visible in the full-resolution and tapered 20~cm maps, on the 
{\it opposite\/} side of the nucleus.  The new northwestern component, located 
approximately 40\asec\ (4.7 kpc) to the northwest at P.A. $\approx$ 110\deg, is 
weaker and more extended than the southeastern component.  The flux densities 
from the tapered 20~cm map are $S_{20}$(SE) = 8.0~mJy and $S_{20}$(NW) = 
3.8~mJy; the northwestern component is not detected at 6~cm.  We believe that 
the northwestern component is real.  Our tapered maps at both frequencies 
(see bottom right panel of Fig.~4{\it c\/}) also show a ``tongue'' of emission 
extending from the central core {\it toward\/} the northwestern component, 
strongly suggestive of an ejection origin for the latter.   The nucleus was not 
detected in the VLBI experiment of Hummel et al. (1982).  We detected weak 
linearly-polarized emission in our data.  The signal is most readily seen in 
naturally-weighted maps tapered to $\Delta \theta$ = 2\farcs5 
(Fig.~16{\it c\/}).  At 6~cm, the signal is roughly equally divided between the 
core ($S^I_{\rm pol, 6}$ = 0.70~mJy) and the southeastern feature 
($S^I_{\rm pol, 6}$ = 0.55~mJy).  The core polarization is less
than 0.2\% of the total-intensity peak, but it appears that
there may be a significant signal in the polarization somewhat
off the location of that peak.  The emission at 20~cm is much less well 
defined; nevertheless, there do appear to be significant clumps of emission 
closely associated with the jet-like extensions in the total-intensity image.  
The individual clumps have peak flux densities $S^P_{\rm pol, 20}\,\approx$ 
0.25~mJy beam$^{-1}$, approximately 10 times the rms in the map.

\noindent {\it NGC~2685 (S2/T2:)}. --- Not detected.  Heckman et al. (1980) 
obtained less stringent upper limits at 6~cm ($S_6\,<$ 2~mJy; $\Delta \theta$
= 1\farcs7) and 20~cm ($S_{20}\,<$ 3~mJy; $\Delta \theta$ = 4\asec).

\noindent {\it NGC~3031, M 81 (S1.5)}. ---  The nucleus of M~81 contains a
bright ($\sim$90~mJy), variable, flat-spectrum radio core (e.g., 
Crane, Giuffrida, \& Carlson 1976; de~Bruyn \etal 1976; Bartel \etal 1982) 
with a one-sided jet on VLBI scales (Bietenholz, Bartel, \& Rupen 2000).  Its 
variability pattern is complex, with occasional outbursts during which the 
source doubles in brightness at higher frequencies (Ho et al. 1999b).  The 
multi-configuration study by Kaufman et al. (1996) reveals a wealth of 
structural details.  We reprocessed their 6~cm B-array and 20~cm A-array 
data, and most of the morphological features discussed by Kaufman et al. 
are recovered in our maps.  The most spectacular feature is an arc-like 
structure located 45\asec\ to the northeast of the nucleus at P.A. $\approx$ 
45\deg, highly suggestive of an outflow origin.  There are, in addition, a 
number of compact sources located within the central $\sim$2\amin\ which are 
likely to be associated with M~81 (see Kaufman et al. 1996 for details).  In 
deriving the largest linear extent of the source, we do not include the two 
compact sources labeled ``57'' and ``75'' in Figure~6 of Kaufman et al. 
(1996); their relation to the nucleus, though suggestive, is uncertain.
Kaufman et al. (1996) have discussed extensively the polarization measurements 
of M~81.  Here, we only remark that we have largely recovered the polarization 
structure found by Kaufman et al., the most notable feature being the high 
degree of polarization along the arc at 20~cm (Fig.~16{\it d\/}).  
At 6~cm, in addition to the weak polarized emission near the core reported 
by Kaufman et al., we also find some polarized signal associated with the arc; 
the polarization fraction approaches 100\% in some locations along the arc.
This very high value could be partly caused by resolving out much of the 
extended total-intensity emission, with the polarization changing on somewhat 
smaller scales to which the interferometers are more sensitive.

\noindent {\it NGC~3079 (S2)}. --- NGC~3079 belongs to a minority of 
edge-on spiral galaxies which exhibit anomalous extended emission along the
minor axis.  Apart from the strong nuclear source, the radio morphology is 
dominated by a striking system of loops and bubbles which strongly 
suggest an outflow origin (Duric \& Seaquist 1988, and references therein).  
Our full-resolution images have resolved out most of the extended emission, 
although some of it is recovered in the tapered map.  The extended, bubble-like 
component lies at P.A. $\approx$ 64\deg, close to the optical minor axis, 
which is at P.A. = 75\deg. The nucleus, with a peak flux density of 90~mJy at 
6~cm and 70~mJy at 20~cm in our 1\asec\ maps, has an angular size less than 
0\farcs05 at 15~GHz (Hummel et al. 1984; Carral, Turner, \& Ho 1990), but it 
was not detected in the VLBI experiment of Hummel et al. (1982).  
Duric \& Seaquist (1988) detected linearly-polarized 6~cm emission over the 
extent of the bubble-like feature and along the major axis of the galaxy's 
disk.  Our higher resolution map is shown in Figure~16{\it g}.  The 
integrated polarized flux density of the bubble-like feature is 
$S^I_{\rm pol, 6}$ = 12.4~mJy, and the corresponding area-averaged 
polarization fraction is very high, $\sim$73\%.  This high percentage 
polarization may be caused by polarization structure on smaller scales than the 
total-intensity structure, as for NGC~3031.  We did not detect 
polarization at 20~cm ($S^P_{\rm pol, 20}\,<$ 0.096~mJy beam$^{-1}$).

\noindent {\it NGC~3147 (S2)}. ---  Several investigators have observed
this galaxy at $\sim$1\asec\ resolution at 6~cm and find a compact, 
$\sim$10~mJy core (Heckman et al. 1980; van der Hulst et al. 1981; Vila et al. 
1990; Laurent-Muehleisen et al. 1997), consistent with our measurements.  
Vila et al. (1990) report $S_{20}$ = 10.6~mJy, also in good agreement with 
our value.  The 6~cm core shows weak, but significant, linear polarization 
(Fig.~16{\it f\/}); the peak signal, $S^P_{\rm pol, 6}$ = 0.23~mJy beam$^{-1}$, 
is approximately 10 times the rms.  Polarization is not detected at 20~cm 
($S^P_{\rm pol, 20}\,<$ 0.11~mJy beam$^{-1}$).

\noindent {\it NGC~3185 (S2:)}. --- Marginally detected at 6~cm but not 
at 20~cm.  Hummel et al (1987) quote an uncertain detection of 0.7~mJy 
at 20~cm ($\Delta \theta$ = 1\farcs3), considerably stronger than our upper
limit of 0.16~mJy at a similar resolution.

\noindent {\it NGC~3227 (S1.5)}. --- The core of NGC~3227 is resolved at 
\lax 1\asec\ resolution (Condon 1980; van der Hulst et al. 1981; Ulvestad et 
al. 1981; Ulvestad \& Wilson 1984a; Hummel et al. 1987; Kukula et al. 1995), 
and at even higher resolution it breaks up into a 0\farcs4 double source at 
P.A. $\approx$ 170\deg\ (Mundell et al. 1995; $\Delta \theta$ = 0\farcs05).

\noindent {\it NGC~3254 (S2)}. --- Not detected. Hummel et al. (1985) obtained 
a less stringent upper limit of $S_{20}\,<$ 1~mJy ($\Delta \theta$ = 15\asec).

\noindent {\it NGC~3486 (S2)}. --- Not detected. Hummel et al. (1987) obtained 
a less stringent upper limit of $S_{20}\,<$ 0.5~mJy ($\Delta \theta$ = 
1\farcs3).

\noindent {\it NGC~3516 (S1.2)}. --- The shape of the synthesized beam 
is highly elongated, and our maps do not sample well the extended emission in 
this object.  A strong (122~mJy at 6~cm, 326~mJy at 20~cm) background source 
4\farcm3 to the southeast 
also contributes sidelobe confusion.   The central core component in NGC~3516 
has been well documented in previous observations at 20, 6, 3.6, and 2~cm 
(Ulvestad \& Wilson 1984b, 1989; Kukula et al. 1995; Nagar et al. 1999). 
The deeper 20~cm map ($\Delta \theta$ = 1\asec) presented by Miyaji, Wilson, 
\& P\'erez-Fournon (1992) shows elongated, one-sided emission extending 
20\asec\ northeast of the nucleus.  The apparent spatial coincidence of the 
radio continuum with optical emission-line nebulosity suggested to these 
authors that the two components are related to a bipolar gaseous outflow 
from the nucleus.  Our 20~cm map clearly detected the same northeast 
extension, but we also see a weaker counterpart, of approximately the same 
length,  toward the southwest.   Both have roughly equal strength, 
each $\sim$4~mJy.  The double-sidedness of the linear feature is even more 
evident in the tapered 6~cm map.  The total angular extent of the source is 
$\sim$51\asec, which corresponds to a projected length of 9.6 kpc, along 
P.A. = 38\deg.  The  southwestern extension can also be seen in the 
$\Delta \theta$ = 6\asec\ 6~cm WSRT image of Baum et al. (1993).  It is of 
interest to note that the \oiii\ \lamb 5007 image of Miyaji et al. (1992) 
shows significant line emission to the southwest of the nucleus, roughly 
coincident with the southwestern jet component.

\noindent {\it NGC~3735 (S2:)}. ---  The central core of NGC~3735 is embedded 
within a much larger disk of radio emission, which is best seen in the tapered 
6~cm map.  The extended component, with an angular diameter of 76\asec\ (15 
kpc), has the same P.A. as that of the major axis of this edge-on galaxy, and 
is most likely associated with it.  Irwin, English, \& Sarathia (1999) show a 
low-resolution C-array 20~cm image of the galaxy.  Although the fit of the 
central core in our maps formally indicates that it is slightly resolved, 
we consider this result to be unreliable because the derived major axis of the
source lies along the very elongated beam, and thus we classify the core 
as unresolved.  Greenhill et al. (1997) reported the discovery of a 
nuclear water maser source in NGC~3735.  They also observed the nucleus in 
the continuum at high resolution ($\Delta \theta\,\approx$ 0\farcs2--0\farcs4).
The nucleus was detected at 3.6~cm ($S_{3.6}$ = 0.6~mJy) and 6~cm ($S_6$ = 
0.7~mJy), but not at 2~cm ($S_2 \,<$  1~mJy).  Their 6~cm flux density is 
consistent with our measurement ($S^P_6$ = 0.81~mJy).

\noindent {\it NGC~3941 (S2:)}. --- Marginally detected at 6~cm but not 
at 20~cm.  Wrobel \& Heeschen (1991) placed an upper limit of 0.5~mJy 
in a 6~cm map made with $\Delta \theta$ = 5\asec.

\noindent {\it NGC~3976 (S2:)}. --- A 0.5~mJy compact core was detected 
at 6~cm at the level of 11.5 $\sigma$ in the full-resolution map and 
9 $\sigma$ in the tapered map.  Yet, the 20~cm maps show no evidence 
of any significant signal, suggesting that the 6~cm source is absorbed at 
lower frequencies.  There are no previously published radio observations of 
this galaxy.

\noindent {\it NGC~3982 (S1.9)}. --- Ulvestad \& Wilson (1989) detected the 
source at 6~cm but not at 20~cm because of strong confusion; there is a 
421~mJy background source in the field (cf. Table~4).  Kukula et al. 
(1995) give a measurement at 3.6~cm.  We detected an unresolved core at 6 and 
20~cm.  In addition, our tapered 6~cm map shows a weak, elongated feature 
roughly 4 kpc in length along P.A. = 0\deg\--10\deg.  The feature is more 
evident toward the south of the nucleus, but it can be seen also 
extending to the north of it.  

\noindent {\it NGC~4051 (S1.2)}. --- The radio structure of the nuclear 
region is known to be quite complex.  At the highest available resolution, 
the central core splits into a small-scale double separated by 0\farcs4 
roughly in the east-west direction (Ulvestad \& Wilson 1984b, 6~cm, $\Delta 
\theta$ = 0\farcs4; Kukula et al. 1995, 3.6~cm, $\Delta \theta$ = 0\farcs3).
On a somewhat coarser resolution of $\sim$1\asec, however, particularly at 
20~cm, the source is dominated by emission extending toward the southwest 
(Ulvestad \& Wilson 1984b; Vila et al. 1990).  We detected this extended 
component in our maps, both at 6 and 20~cm, but we additionally see a fainter,
oppositely-directed component to the northeast.  The two-sidedness of the 
extended component is best viewed in our tapered 6~cm map; it can also be seen 
in the $\Delta\theta\,\approx$ 4\asec\ 6~cm WSRT map of Baum et al. (1993) .  
The total linear extent of the ``jet'' is 1.2 kpc, and it lies along P.A. = 
41\deg, perpendicular to the host-galaxy major axis (P.A. = 135\deg).  We 
note, in passing, that the nucleus appears to be quite variable at 6~cm.  The 
existing measurements made with $\Delta \theta\,\approx$ 1\asec\ 
show a spread of a factor of 6 in the 6~cm flux density (van der Hulst et al. 
1981, 3.5~mJy; Ulvestad \& Wilson 1984b, 6.0~mJy; Vila et al. 1990, 0.9~mJy;
this paper, 3.2~mJy).

\noindent {\it NGC~4138 (S1.9)}. --- The nucleus is badly confused by two very 
strong sources ($S_{20}$ = 1.3 and 0.33~Jy) located $\sim$3\farcm 5 to the 
southwest.  The rms noise of the tapered 20~cm map is quite high.  
Nonetheless, a weak, compact core with an inverted spectrum 
($S_6$ = 0.78~mJy, $S_{20}$ = 0.51~mJy, $\alpha_6^{20}$ = 0.45) is seen 
at the position of the optical nucleus.  The only other previous 
radio observation is that of Wrobel \& Heeschen (1991), who placed an 
upper limit of $S_6\,<$ 2~mJy ($\Delta \theta$ = 5\asec).

\noindent {\it NGC~4151 (S1.5)}. --- The flux densities we measured are in 
good agreement with those of previous arcsecond-scale observations 
(e.g., Ulvestad et al. 1981, $S_6$ = 125~mJy), but our maps have insufficient 
resolution to reveal the intricate fine structure of this complex radio 
source.  Sub-arcsecond VLA maps have been published by Carral et al. (1990; 
2~cm), Kukula et al. (1995; 3.6~cm), and Pedlar et al. (1993; 3.6 and 6~cm), 
and a sub-arcsecond MERLIN image has been published by Booler, Pedlar, \& 
Davies (1982; 18~cm). Ulvestad et al. (1998) presented VLBA 6 and 18~cm maps 
and an interpretation of the various radio components.  Although this 
source has previously been extensively studied in the radio, we are not aware
of any polarization measurements.  We detected linearly-polarized emission 
both at 6 and 20~cm (Fig.~16{\it h\/}).  The 6~cm emission appears slightly 
resolved ($S^I_{\rm pol, 6}$ = 0.56~mJy), with an extension to the west of the 
nucleus.  At 20~cm the emission is concentrated in an unresolved clump 
($S^P_{\rm pol, 20}$ = 0.31~mJy beam$^{-1}$) which is slightly displaced 
to the southwest of the nucleus.

\noindent {\it NGC~4168 (S1.9:)}. --- Our 6~cm flux density of 5.2~mJy for 
the compact source agrees well with the $\Delta \theta$ = 5\asec\ measurement 
of Wrobel \& Heeschen (1991), who give $S_6$ = 4.5~mJy.  
%Turner, Helou, \& Terzian (1988) observed NGC~4168 using the two-element 
%Arecibo interferometer, and they measured $S$ = 7.9~mJy at 2.4~GHz.  This 
%value is significantly higher than expected from interpolation of our 
%6 and 20~cm points, suggesting that the nucleus is variable.  

\noindent {\it NGC~4169 (S2)}. ---  A very weak ($S^P_6$ = 0.19~mJy, 
$S^P_{20}$ = 0.21~mJy) source consistently appears slightly offset to the 
northwest of the optical position of the nucleus.  This is seen also in the 
tapered 6 and 20~cm maps.  Although the detection is somewhat marginal, it is 
probably real, especially since a 1.5~mJy source was detected in the 
lower resolution FIRST survey (Becker, White, \& Helfand 1995; 20~cm, 
$\Delta \theta$ = 5\asec).  The position of the FIRST source agrees 
with our radio position.

\noindent {\it NGC~4203 (L1.9)}. --- The nucleus of this galaxy is classified
as a LINER~1.9 and is not formally part of our Seyfert sample.  Most of the 
radio flux is contained in a central, unresolved core, whose spectrum 
is inverted ($\alpha_6^{20}$ = 0.44).  Our 6~cm flux density of 11.2~mJy is 
in excellent agreement with those from previous studies (Fabbiano, Gioia, \& 
Trinchieri 1989, $S_6$ = 11.6~mJy, $\Delta \theta\, \approx$ 1\asec; Wrobel \& 
Heeschen 1991, $S_6$ = 12.5~mJy, $\Delta \theta$ =  5\asec), and our 20~cm 
measurement of 6.5~mJy matches well the value from the FIRST survey ($S_{20}$ 
= 6.9~mJy).   The inverted spectrum, accompanied by an apparent lack of 
variability, may indicate free-free absorption rather than synchrotron 
self-absorption.

\noindent {\it NGC~4235 (S1.2)}. --- The unresolved core has a flat or 
slightly inverted spectrum ($\alpha_6^{20}$ = 0.09), consistent with 
previously reported flux densities from 2 to 20~cm (Ulvestad \& Wilson 
1984b, 1989; Kukula et al. 1995).

\noindent {\it NGC~4258, M 106 (S1.9)}. --- At the scale of our images, 
the nucleus of NGC~4258 appears as a slightly resolved 2--3~mJy core 
with a somewhat flat spectral index of $\alpha_6^{20}$ = --0.18, in agreement 
with previous observations from 2 to 20~cm at similar resolution (Vila et al. 
1990; Saikia et al. 1994; Turner \& Ho 1994).  Our 20~cm map shows 
low-surface brightness emission extending to the northwest and southeast.  
These belong to the complex set of ``anomalous radio arms'' best seen in the 
$\Delta \theta$ = 6\farcs5 20~cm VLA+WSRT image of van Albada \& van der Hulst 
(1982).  On VLBI scales, the central core breaks up into a sub-parsec jet 
oriented along the axis of the water maser disk (Herrnstein et al. 1997, 
1998).  The continuum source at 22~GHz varies by a factor of 2 on a timescale 
of weeks (Herrnstein et al. 1997).

\noindent {\it NGC~4378 (S2)}. --- This galaxy was marginally detected at 
6~cm but not at 20~cm. There are no previously published radio observations.

\noindent {\it NGC~4388 (S1.9)}. --- Our maps show a strong central source 
which resolves into two peaks separated by $\sim$1\farcs9 (150 pc) along 
P.A. $\approx$ 15\deg.  Neither peak coincides with the position of the 
optical nucleus.  A long ($\sim$15\asec\ or 1.2 kpc) plume of emission 
extends to the northeast, at the end of which it appears to bend at abruptly.
In the tapered 6~cm map we find an additional ``tongue'' of emission, 
$\sim$20\asec\ long, extending east of the central source.  We fitted the 
central source with a two-Gaussian model.  The northern peak (labeled 
``core-N'' in Table~3) is brighter than the southern one (``core-S'') at 6~cm, 
but the situation is reversed at 20~cm.  The spectral indices for the northern 
and southern peak are, respectively, $\alpha_6^{20}$ = --0.32 and --0.80.  The 
overall morphology and flux densities derived from our maps agree very well 
with previous VLA studies conducted at similar resolutions (Stone, Wilson, 
\& Ward 1988; Hummel \& Saikia 1991).  Higher resolution maps of NGC~4388 have 
been made at 2~cm (Carral et al. 1990), 3.6~cm (Kukula et al. 1995; Falcke, 
Wilson, \& Simpson 1998; Mundell et al. 2000) and 6~cm (Mundell et al. 2000).  
These studies collectively show that the northern peak has a flat,
or perhaps even slightly inverted, spectrum up to 2~cm, and that it is 
unresolved with a size upper limit of 70 mas. These characteristics 
suggest that the northern peak marks the true location of the nucleus, 
which is obscured at optical wavelengths.

\noindent {\it NGC~4395 (S1.8)}. --- NGC~4395 holds the distinction of hosting 
the Seyfert~1 nucleus with the lowest known optical luminosity (Filippenko 
\& Sargent 1989), and one of the intrinsically weakest nuclear X-ray sources 
observed so far (Lira et al. 1999; Moran et al. 1999).  NGC~4395 is also 
highly unusual as an AGN host because of its late Hubble type: it is 
classified as a Magellanic spiral (Sm), only 3.6 Mpc away, with an absolute 
magnitude of $M_{B_T}^0 $ = --17.2.  Sramek (1992; see also Moran et al. 1999) 
previously observed the nucleus with the VLA, and he detected an unresolved 
core with $S_6$ = 0.6~mJy ($\Delta \theta$ = 0\farcs4) and $S_{20}$ = 1.24~mJy 
($\Delta \theta$ = 1\farcs4).  We measure $S_{6}$ = 0.80~mJy and $S_{20}$ = 
1.68~mJy, which indicates that the source has not varied significantly.  Our 
6~cm map is more sensitive than Sramek's, who made use of archival data taken 
in 1982. 

\noindent {\it NGC~4450 (L1.9)}. ---  The nucleus of this galaxy is classified
as a LINER~1.9 and is not formally part of our Seyfert sample.  We detected 
a central, unresolved core, with a flat or mildly inverted spectrum 
($\alpha_6^{20}$ = 0.07).  The previously reported measurement at 20~cm 
(Hummel et al. 1987; $S_{20}$ = 5.3~mJy, $\Delta \theta$ = 1\farcs3) agrees 
reasonably well with ours.

\noindent {\it NGC~4472, M 49 (S2::)}. --- Previous low-resolution maps 
find the radio core to be straddled by double-sided jets with an extent of 
$\sim$2\amin\ (Ekers \& Kotanyi 1978; Condon \& Broderick 1988).  Our maps are 
insensitive to large-scale structure of this angular size, and we detect mainly 
an extended, elliptical source of length $\sim$11\asec\ (9 kpc) along P.A. 
$\approx$ 80\deg.  Van der Hulst et al. (1981) and Fabbiano et al. (1989) 
have published 6~cm observations at similar resolution as ours.  Birkinshaw 
\& Davies (1985) quote an unpublished higher resolution map by R. Laing 
which shows that the central source resolves into a 3\asec\ jet 
along P.A. = 83\deg.  We have detected linearly-polarized emission at 
6 ($S^I_{\rm pol, 6}$ = 1.26~mJy) and 20~cm ($S^I_{\rm pol, 20}$ = 0.40~mJy). 
At both frequencies the strongest polarization lies offset from the 
peak in the total-intensity image (Fig.~16{\it j\/}); the polarization vectors 
appear to follow smaller scale structure which is not well resolved in our maps.

\noindent {\it NGC~4477 (S2)}. --- This galaxy was marginally detected at 6~cm 
but not at 20~cm.  The very faint source ($S_6$ = 0.18~mJy in the untapered map 
and 0.26~mJy in the tapered map) is consistent with the published 
upper limit of $S_6\,<$ 0.25~mJy ($\Delta \theta\,\approx$ 8\asec) by Fabbiano 
et al. (1989) and $S_6\,<$ 0.5~mJy ($\Delta \theta$ = 5\asec) by Wrobel \& 
Heeschen (1991).

\noindent {\it NGC~4501, M 88 (S2)}. ---  We detected an unresolved core in 
both bands.  Hummel et al. (1987) give a 20~cm flux density of $S_{20}$ = 2.7 
mJy ($\Delta \theta$ = 1\farcs3), in reasonable agreement with our value.   
The galaxy was not detected at higher resolution ($\Delta \theta$ = 0\farcs3) 
by Thean et al. (2000) at 3.6~cm; their upper limit of $S_{3.6}\,<$ 0.2~mJy 
implies that the source is variable, it is somewhat resolved at the higher 
resolution, or its spectrum is much steeper than we inferred from our data 
($\alpha_6^{20}$ = --0.48).

\noindent {\it NGC~4565 (S1.9)}. --- VLA flux densities measured with
$\sim$1\asec\ resolution have been published by van der Hulst et al. (1981) 
and Hummel et al. (1987), and a $\Delta \theta$ = 5\asec\ 20~cm detection is 
available from the FIRST survey.  There is a spread of about a factor of 2 
among the various measurements, including ours, suggesting that the nucleus 
may be variable.

\noindent {\it NGC~4579, M 58 (S1.9/L1.9)}. --- The previous 1\asec-resolution 
6~cm map of van der Hulst et al. (1981) reported that the core of 
NGC~4579 is slightly extended along P.A. = 134\deg.  Our maps, especially that 
at 20~cm (see also Hummel et al. 1987), show that the source is a double, 
separated by 4\farcs8 along P.A. = 133\deg.  We identify the brighter of the 
two with the nucleus, since it lies within 0\farcs6 of the optical position of 
the nucleus, which is known to an accuracy of $\pm$0\farcs16.  Our data 
indicate that the core has an inverted spectrum ($\alpha_6^{20}$ = 0.49), 
consistent with the measurements of Sadler et al. (1995) performed with the
Parkes-Tidbinbilla Interferometer at 2.3 and 8.4~GHz ($\alpha$ = 0.19).  Other 
literature data include a VLBI detection at 6~cm ($S$ = 21~mJy, quoted in 
Hummel et al. 1987), an Arecibo interferometer measurement at 2.4~GHz 
($\Delta \theta$ = 1\asec, $S$ = 20~mJy; Turner, Helou, \& Terzian 1988), and 
a VLA point at 3.6~cm ($\Delta \theta$ = 0\farcs3, $S$ = 36.5~mJy; Thean et 
al. 2000).

\noindent {\it NGC~4639 (S1.0)}. --- This galaxy was marginally detected at 
6~cm but not at 20~cm.  Hummel et al. (1985) obtained a less stringent upper 
limit of $S_{20}\,<$ 1~mJy ($\Delta \theta$ = 15\asec).

\noindent {\it NGC~4698 (S2)}. --- This galaxy was marginally detected at
6~cm but not at 20~cm.  Hummel et al. (1987) obtained a less stringent upper 
limit  of $S_{20}\,<$ 1.5~mJy ($\Delta \theta$ = 1\farcs3).

\noindent {\it NGC~4725 (S2:)}. --- Not detected.  No previous 
high-resolution radio observations.

\noindent {\it NGC~5033 (S1.5)}. --- Previous 6 and 20~cm VLA data show
an unresolved (Ulvestad \& Wilson 1989; Collison et al. 1994; Kukula et al. 
1995) or slightly resolved (van der Hulst et al. 1981) core.  Kukula et al. 
remark that their C-array 3.6~cm map contains extended emission on a scale of 
30\asec.  Our 1\asec\ maps show a slightly resolved, steep-spectrum 
($\alpha_6^{20}$ = --0.52) core surrounded by a fluffy envelope with a 
major-axis diameter of $\sim$10\asec\ (0.9 kpc), roughly along the east-west 
direction.  The extended emission is more prominent toward the east.  The 
tapered maps further show a spectacular ridge of emission extended over 
$\sim$40\asec\ (3.6 kpc), running nearly north-south, along the galaxy major 
axis (P.A. = 170\deg).  This large-scale feature lies roughly orthogonal to 
the 10\asec\ structure.

\noindent {\it NGC~5194, M 51 (S2)}. ---  Ford et al. (1985) first 
drew attention to the morphological details of the radio emission in the 
center of this galaxy.  In addition to the ``extra-nuclear cloud'' located 
$\sim$4\asec\ to the southeast of the nucleus, they noticed the ring-like 
appearance of the diffuse emission to the northwest of the nucleus.  Ford et al. 
postulated that these features can plausibly be interpreted as bubbles created 
by a bipolar jet or outflow emanating from the nucleus.  A much deeper, 
sub-arcsecond 6~cm map published by Crane \& van der Hulst (1992) shows 
convincingly that the southeastern lobe of emission is connected to the 
nucleus by a thin, sinuous, jet-like feature, and the overall radio structure 
is embedded within the extended optical emission-line region imaged by 
Cecil (1988).  Our maps qualitatively resemble the earlier versions.  We wish 
only to draw attention to a knot of emission which lies $\sim$28\asec\ (1 kpc) 
to the northwest of the nucleus, roughly along a straight line 
connecting the other features.  The source has $S_{20}$ = 1.9~mJy and 
$S_6$ = 1.4~mJy, and it is unresolved.  Since the density of background 
sources stronger than $S_{20}\,\approx$ 2~mJy is $\sim$10$^5$ sr$^{-1}$ 
(Windhorst et al. 1985), the probability that the source is unrelated to 
M~51 is quite small ($<$1\%).   It is unclear, however, whether the source is 
directly associated with the outflow itself, in which case the total extent of 
the radio source would be 35\asec\ (1.3 kpc) instead of 24\asec\ (0.9 kpc) as 
we have adopted in Table~5.  Our flux densities for the nucleus are 
consistent with the values of Hummel et al. (1987; $S_{20}$ = 3.6~mJy, $\Delta 
\theta$ = 1\farcs3), Crane \& van der Hulst (1992; $S_6$ = 0.89~mJy, $\Delta 
\theta$ = 0\farcs3), and Turner \& Ho (1994; $S_6$ = 1.1~mJy, $\Delta \theta$ 
= 1\farcs1).  Thean et al. (2000) give $S_{3.6}$ = 0.5~mJy ($\Delta \theta$ = 
0\farcs3).  Crane \& van der Hulst (1992) remarked that their 6~cm data 
show marginally significant linearly-polarized emission coincident with the 
extra-nuclear cloud to the southeast.  Our maps confirm this finding, both at 
6 and 20~cm (Fig.~16{\it i\/}); in addition, a number of other peaks 
($S^P_{\rm pol, 6}\,\approx\,S^P_{\rm pol, 20}\,\approx$ 0.2~mJy beam$^{-1}$) 
of comparable statistical significance appear to be associated with 
ring-like feature to the northwest of the nucleus.

\noindent {\it NGC~5273 (S1.5)}. --- We detect a compact, unresolved core 
at both frequencies.  The source is slightly resolved along 
P.A. = 0\deg\ in the 6~cm data of Ulvestad \& Wilson (1984b).  Both 
Kukula et al. (1995) and Nagar et al. (1999) have observed NGC~5273 at 
3.6~cm with $\Delta \theta$ = 0\farcs3; the former did not detect the 
nucleus ($S_{3.6}\,<$ 0.23~mJy), but the latter did ($S_{3.6}$ = 0.6~mJy), 
which suggests that the source is variable.  

\noindent {\it NGC~5395 (S2/L2)}. --- Not detected.  Hummel et al. (1987)
obtained a less stringent upper limit of $S_{20}\,<$ 0.4~mJy 
($\Delta \theta$ = 1\farcs3).

\noindent {\it NGC~5548 (S1.5)}. --- The classic triple linear structure of 
NGC~5548, consisting of a compact core and oppositely-directed lobes, 
is well known (Wilson \& Ulvestad 1982a; Kukula et al. 1995; Nagar et al. 
1999; Wrobel 2000).  Our maps are consistent with those of previous studies.  
The core is unresolved at our resolution, and the total extent of the jets 
is $\sim$15\asec\ (5 kpc) along P.A. = 163\deg.   We find evidence for 
weak ($S^P_{\rm pol, 6}\,\approx\,S^P_{\rm pol, 20}\,\approx$ 0.2~mJy 
beam$^{-1}$) linearly-polarized emission associated with the extended 
emission (Fig.~16{\it k\/}).  

\noindent {\it NGC~5631 (S2/L2:)}. --- Although this source is quite weak, 
we believe it to be real because it was detected in both bands at 
full-resolution and in the tapered 6~cm map.  The only previous radio 
observation is that of Wrobel \& Heeschen (1991), who obtained an upper 
limit of $S_{6}\,<$ 0.5~mJy with $\Delta \theta$ = 5\asec.  Our 
measurements are consistent with this upper limit.

\noindent {\it NGC~6951 (S2)}. --- NGC~6951 harbors a Seyfert 2 nucleus 
which is surrounded by a 1 kpc-diameter circumnuclear starburst ring (Barth 
et al. 1995).  Our full-resolution maps depict the 
ring with much greater clarity than in previous radio maps made with 
similar resolution (Vila et al. 1990; Saikia et al. 1994).  The nucleus, 
which contains only $\sim$10\% of the total flux in either band, has a 
steep spectrum ($\alpha_6^{20}$ = --0.90).  Although it appears to 
be slightly resolved in one direction, the apparent elongation is 
along the direction of the beam, and we consider the source to be unresolved.

\noindent {\it NGC~7479 (S1.9)}. --- Condon, Frayer, \& Broderick (1991) and 
Hummel et al. (1987) have previously detected a slightly resolved $\sim$6~mJy 
core at 20~cm based on $\Delta \theta$ = 1\farcs5 maps.  We find the core 
to be unresolved at our resolution, and our full-resolution 20~cm and tapered 
6 and 20~cm maps show an additional curved extension to the north, and perhaps 
a weaker, but shorter, component to the south.  The total extent of the linear 
feature is 36\asec, or 5.7 kpc.  We find evidence for
weak ($S^P_{\rm pol, 6}\,\approx\,S^P_{\rm pol, 20}\,\approx$ 0.1--0.2~mJy
beam$^{-1}$) linearly-polarized emission associated with the extended
emission (Fig.~16{\it l\/}).

\noindent {\it NGC~7743 (S2)}. ---  This galaxy has been observed with the 
C-array at 6~cm by Wrobel \& Heeschen (1991) and with the A-array at 3.6
and 20~cm by Nagar et al. (1999).  We reprocessed the archival 6 and 20~cm 
data, and the flux densities of the compact source measured from our maps are 
in good agreement with the published values.  In order to estimate the 
spectral index, we tapered the 20~cm A-array map to a resolution comparable 
to that of the full-resolution 6~cm C-array map ($\Delta \theta\,\approx$ 
3\farcs8); we measure $\alpha_6^{20}$ = --0.49.

%%REFERENCES
%\clearpage
%
%\vskip 0.75truein
%\centerline{\bf{References}}
%\medskip
%

%FIGURE CAPTIONS
\clearpage
\centerline{FIGURE CAPTIONS}
\bigskip

{\it Fig. 1. ---} Number distributions of ({\it a\/}) distance, ({\it b\/}) 
total absolute blue magnitude (corrected for internal extinction), ({\it c\/}) 
morphological type index, $T$ (--3 to --1 = S0,  1 = Sa, 3 = Sb, 5 = Sc, 7 = 
Sd,  9 = Sm), and ({\it d\/}) extinction-corrected luminosity of the narrow 
H\al\ emission line for the Palomar Seyfert sample.  The morphologically 
peculiar galaxy NGC~1275 was omitted from ({\it c\/}).  The type~1 objects are 
shown as shaded histograms, and the total sample (types~1 and 2) is denoted by 
the unshaded histograms.  All the data are taken from Ho et al. (1997a) and 
are given in Tables~1 and 5.
 
\bigskip
{\it Fig. 2. ---} Full-resolution ($\Delta \theta\,\approx$ 1\asec), 
uniformly-weighted maps at 20~cm ({\it a\/}) and 6~cm ({\it b\/}), and tapered, 
naturally-weighted maps at 6~cm ({\it c\/}) for NGC~ 185, NGC~ 676, NGC~ 777, 
and NGC~1058.  The resolution of the tapered maps, given in Table~3, is
either $\Delta \theta\,\approx$ 2\farcs5 or 3\farcs6.  Contour plots have been 
overlaid on the greyscale images to improve the visibility of the full dynamic 
range of the maps.  Panels ({\it a\/}) and ({\it b\/}) are registered, and they 
have the same dimensions (30\asec $\times$ 30\asec).  The dimensions of panel 
({\it c\/}) have been optimized for each galaxy.  The restoring beam is 
depicted as a hatched ellipse on the lower left-hand corner of each map.   The 
maps display contour levels of rms $\times$ (--6, --3, 3, 6, 12, 24, 48, ...); 
the rms values for the maps are listed in Table~3.   The optical position of 
the galaxy is marked with a cross, the semi-major length of which corresponds 
to the 1 $\sigma$ uncertainties given in Table~1.  

{\it Fig. 3. ---} Maps for NGC~1068, NGC~1167, NGC~1275, and NGC~1358.  Same 
as in Fig. 2.  
\bigskip

{\it Fig. 4. ---} Maps for NGC~1667, NGC~2273, NGC~2639, and NGC~2655.  Same 
as in Fig. 2.  A tapered 20~cm map of NGC~2655 is also shown as the top image 
in panel ({\it c\/}). 
\bigskip

{\it Fig. 5. ---} Maps for NGC~2685, NGC~3031, NGC~3079, and NGC~3147.   Same 
as in Fig. 2, except that the dimensions of panels ({\it a\/}) and ({\it b\/}) 
for NGC~3079 are 48\asec $\times$ 48\asec.
\bigskip

{\it Fig. 6. ---} Maps for NGC~3185, NGC~3227, NGC~3254, and NGC~3486.   Same 
as in Fig. 2.
\bigskip

{\it Fig. 7. ---} Maps for NGC~3516, NGC~3735, NGC~3941, and NGC~3976.  Same 
as in Fig. 2.
\bigskip

{\it Fig. 8. ---} Maps for NGC~3982, NGC~4051, NGC~4138, and NGC~4151.   Same 
as in Fig. 2.
\bigskip

{\it Fig. 9. ---} Maps for NGC~4168, NGC~4169, NGC~4203, and NGC~4235.   Same 
as in Fig. 2.
\bigskip

{\it Fig. 10. ---} Maps for NGC~4258, NGC~4378, NGC~4388, and NGC~4395.  Same 
as in Fig. 2.
\bigskip

{\it Fig. 11. ---} Maps for NGC~4450, NGC~4472, NGC~4477, and NGC~4501.   Same 
as in Fig. 2.
\bigskip

{\it Fig. 12. ---} Maps for NGC~4565, NGC~4579, NGC~4639, and NGC~4698.   Same 
as in Fig. 2.
\bigskip

{\it Fig. 13. ---} Maps for NGC~4725, NGC~5033, NGC~5194, and NGC~5273.  Same 
as in Fig. 2, except that the dimensions of panels ({\it a\/}) and ({\it b\/}) 
for NGC~5194 are 40\asec $\times$ 40\asec.
\bigskip

{\it Fig. 14. ---} Maps for NGC~5395, NGC~5548, NGC~5631, and NGC~6951.   Same 
as in Fig. 2.
\bigskip

{\it Fig. 15. ---} Maps for NGC~7479 and NGC~7743.   Same as in Fig. 2.  
The 6~cm images of NGC~7743 were made with the C-configuration rather than 
the B-configuration, and thus have poorer resolution than the other images.
\bigskip

{\it Fig. 16. ---} Objects with detected linear polarization.   The 
contours show maps of total intensity, as in Fig. 2.  The maps are 
uniformly weighted and have full resolution ($\Delta \theta\,\approx$ 1\asec).
The vectors represent linearly-polarized emission derived from 
naturally-weighted maps with $\Delta \theta\,\approx$ 1\farcs4; the
position angles of the vectors denote the direction of the electric field.
Only signal in excess of 3 times the rms is shown.   For NGC~2655, NGC~3031, 
and NGC~7479, both the total-intensity and linearly-polarized maps are
naturally weighted and tapered to $\Delta \theta\,\approx$ 2\farcs5.  
A polarization vector 1\asec\ long represents the following 
linearly-polarized flux density:
({\it a\/})    1~mJy beam$^{-1}$, NGC~1068; 
({\it b\/})    1~mJy beam$^{-1}$, NGC~1167; 
({\it c\/})  0.1~mJy beam$^{-1}$, NGC~2655; 
({\it d\/})  0.1~mJy beam$^{-1}$, NGC~3031; 
({\it e\/})  0.5~mJy beam$^{-1}$, NGC~2639; 
({\it f\/})  0.5~mJy beam$^{-1}$, NGC~3147; 
({\it g\/}) 0.25~mJy beam$^{-1}$, NGC~3079; 
({\it h\/}) 0.25~mJy beam$^{-1}$, NGC~4151; 
({\it i\/}) 0.25~mJy beam$^{-1}$, NGC~5194; 
({\it j\/})  0.5~mJy beam$^{-1}$, NGC~4472; 
({\it k\/}) 0.25~mJy beam$^{-1}$, NGC~5548; and
({\it l\/})  0.1~mJy beam$^{-1}$, NGC~7479.

{\it Fig. 17. ---}  
Number distributions of radio powers.  The panels show, from top to
bottom, measurements for the total integrated power at 6~cm, the peak core 
power at 6~cm, the total integrated power at 20~cm, and the peak core 
power at 20~cm.  The shaded regions of the histogram denote upper limits.
\bigskip

{\it Fig. 18. ---} 
Number distributions of ({\it a\/}) linear source size and ({\it b\/})
spectral index between 6 and 20~cm.
The shaded regions of the histogram denote upper limits.

%\clearpage
 
%TABLES
%Table 1 has bounding box = 50 50 590 730
%\clearpage
\begin{figure}
%\plotone{table1_v4.ps}
\hskip 0.5in
\psfig{file=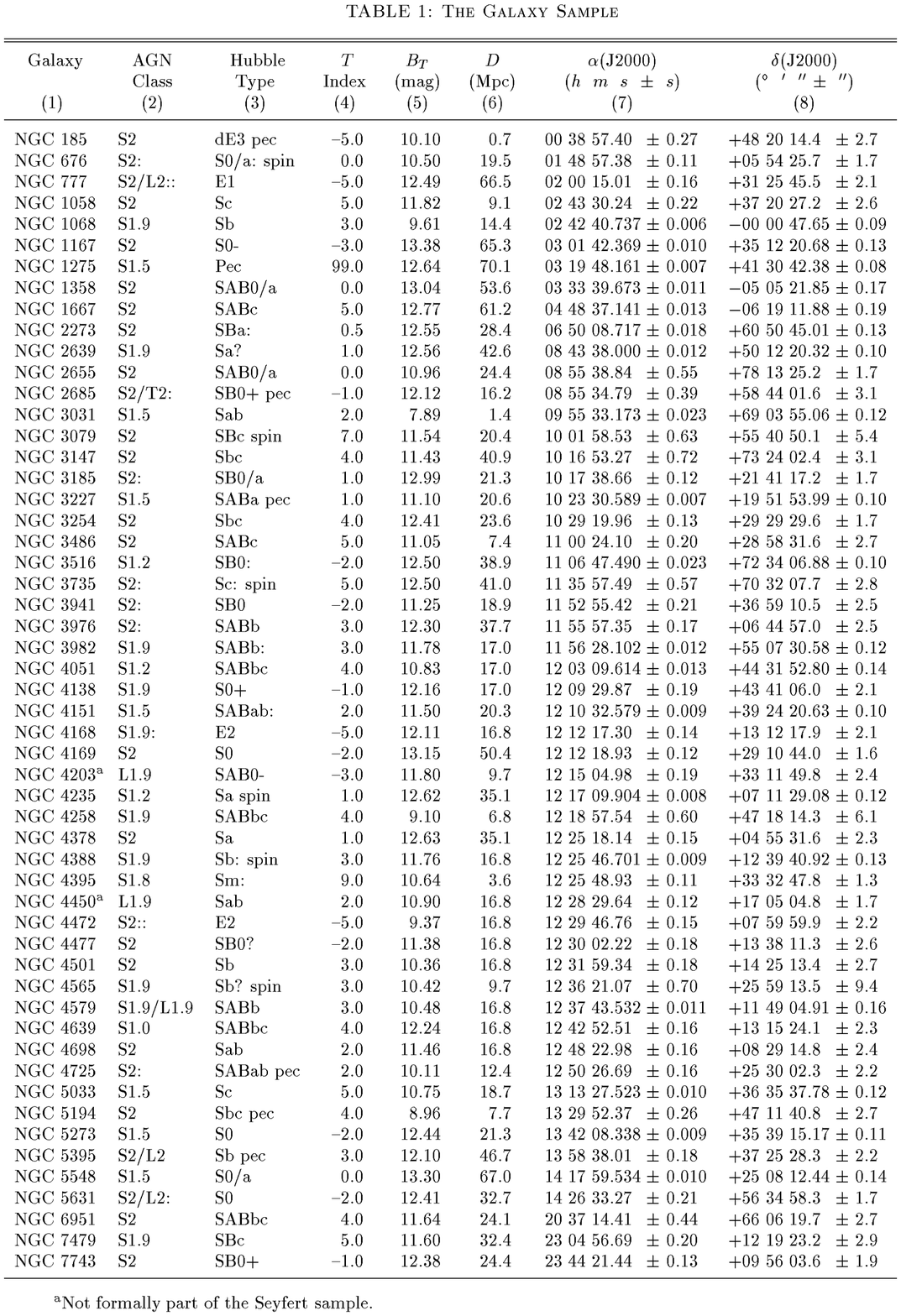,height=8.5in}
\end{figure}

%Table 2 has bounding box = 50 50 500 500
%\clearpage
\begin{figure}
\plotone{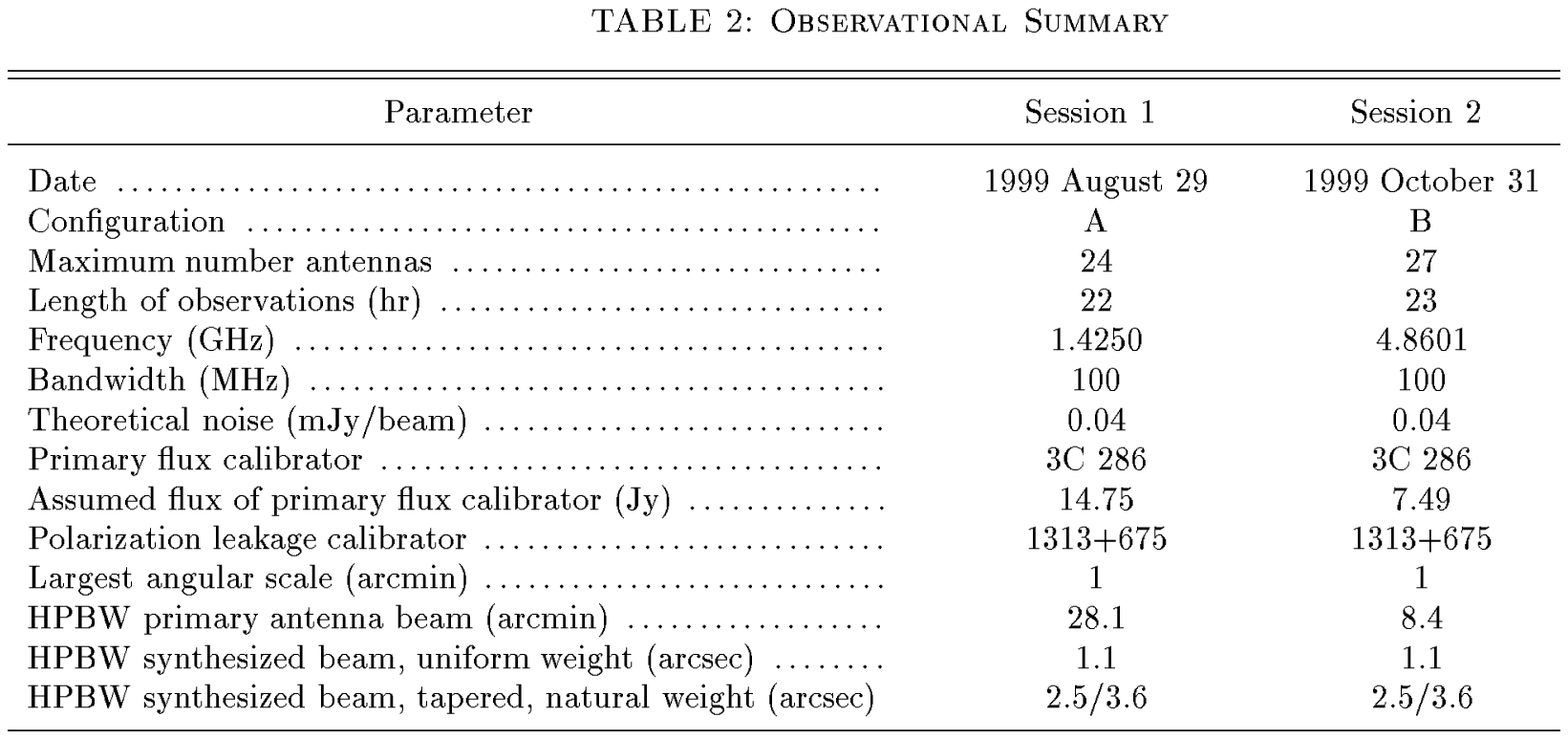}
\end{figure}

%\clearpage
%Table 3 has bounding box = 50 50 500 750 (change to Portrait in ps)
\psfig{file=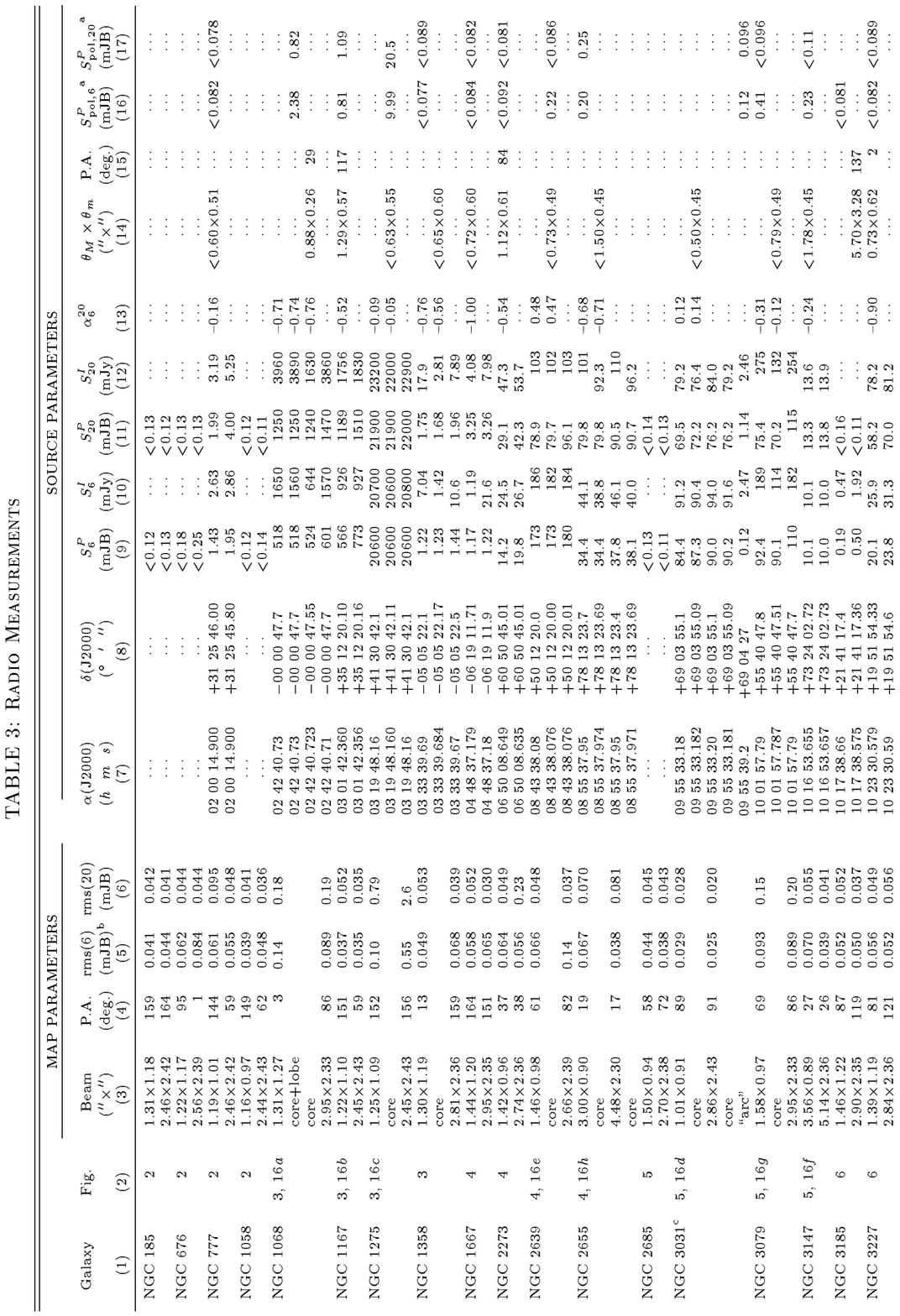,height=8.5in,width=7.0in,angle=180}

%\clearpage
%bounding box = 50 50 590 750 (change to Portrait in ps)
\psfig{file=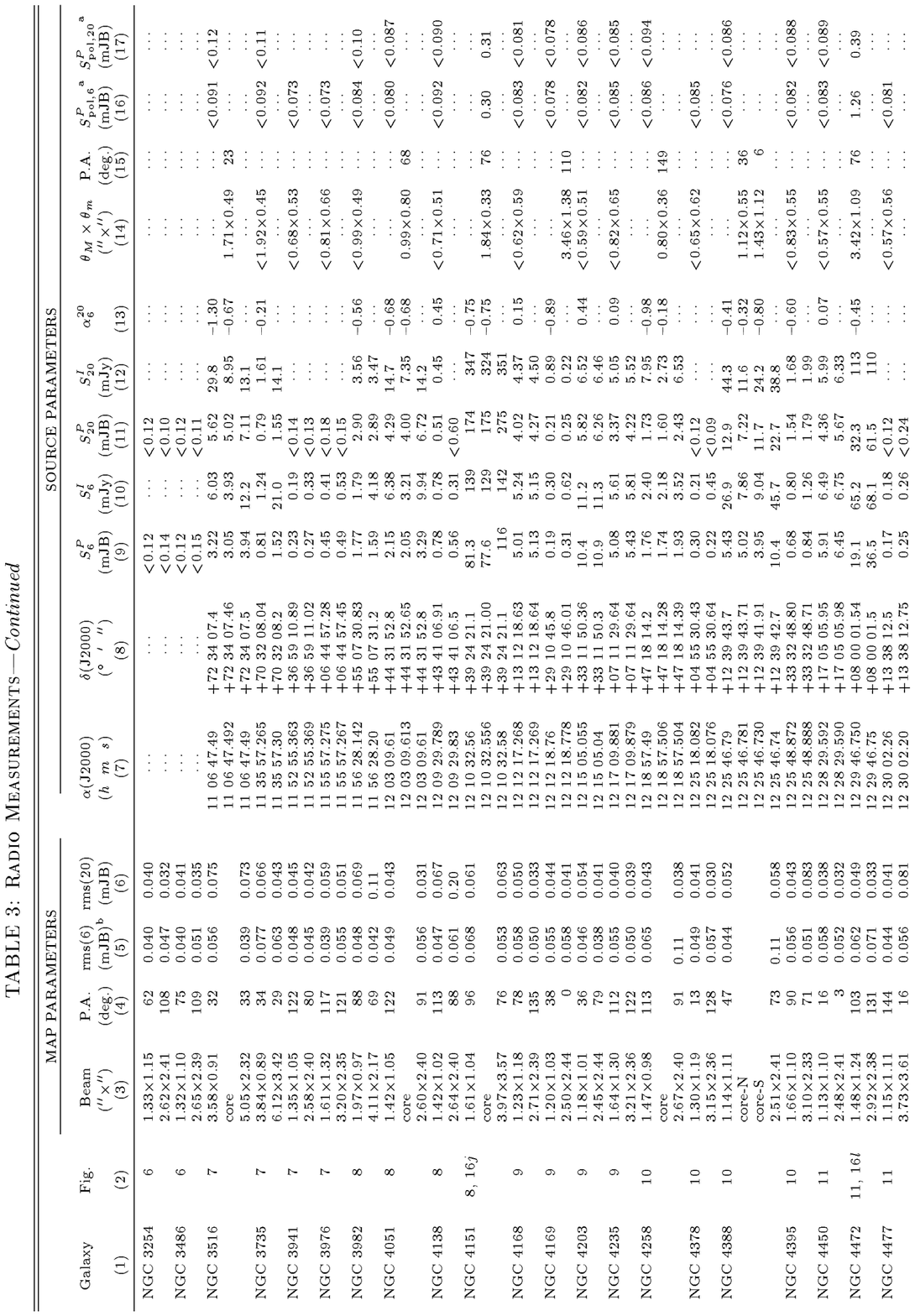,height=8.5in,width=7.0in,angle=180}

%\clearpage
%bounding box = 50 70 500 780 (change to Portrait in ps)
\psfig{file=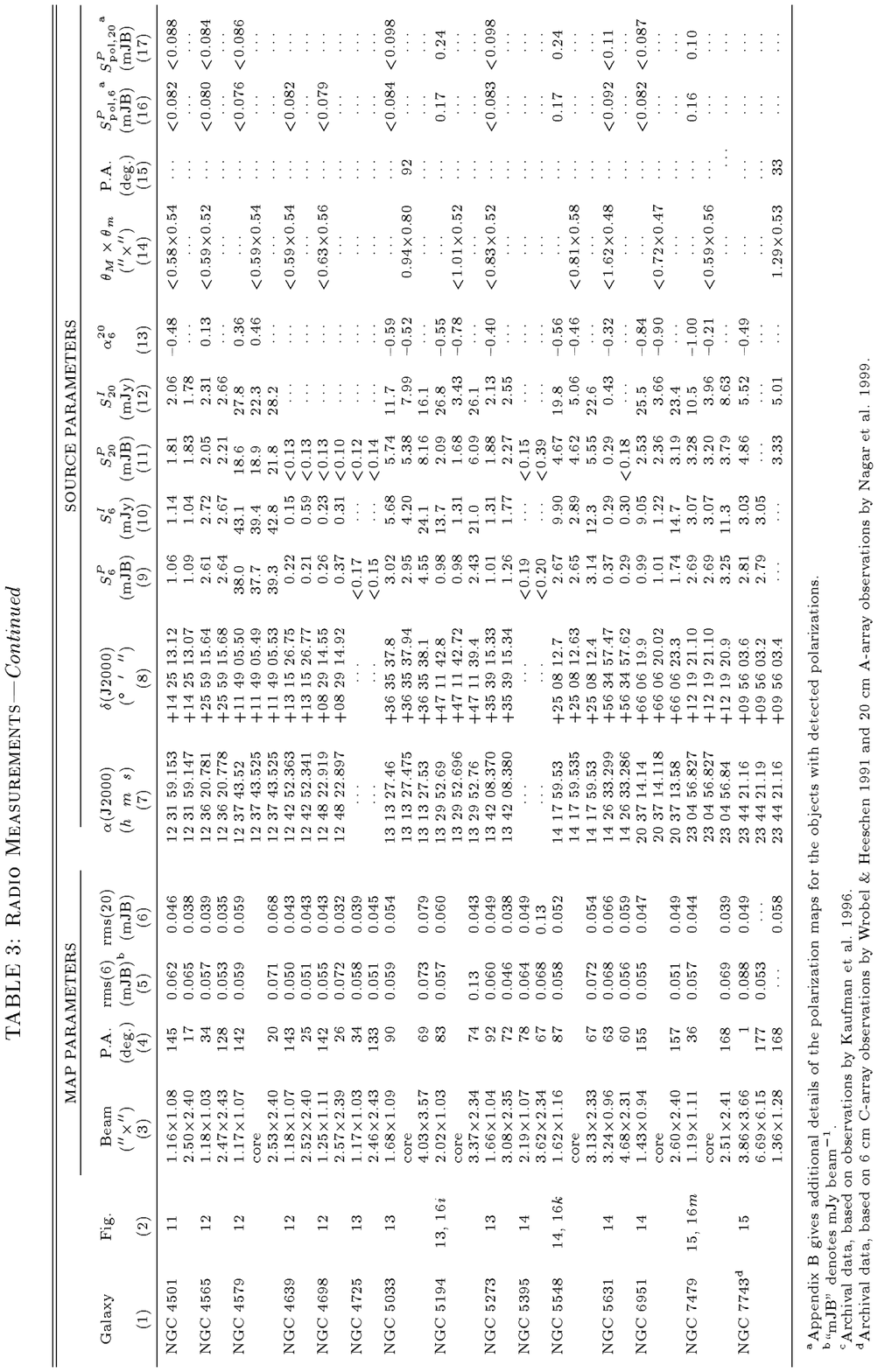,height=8.5in,width=7.0in,angle=180}

%Table 4 has bounding box = 50 90 590 670
%\clearpage
\hskip 0.2in
\psfig{file=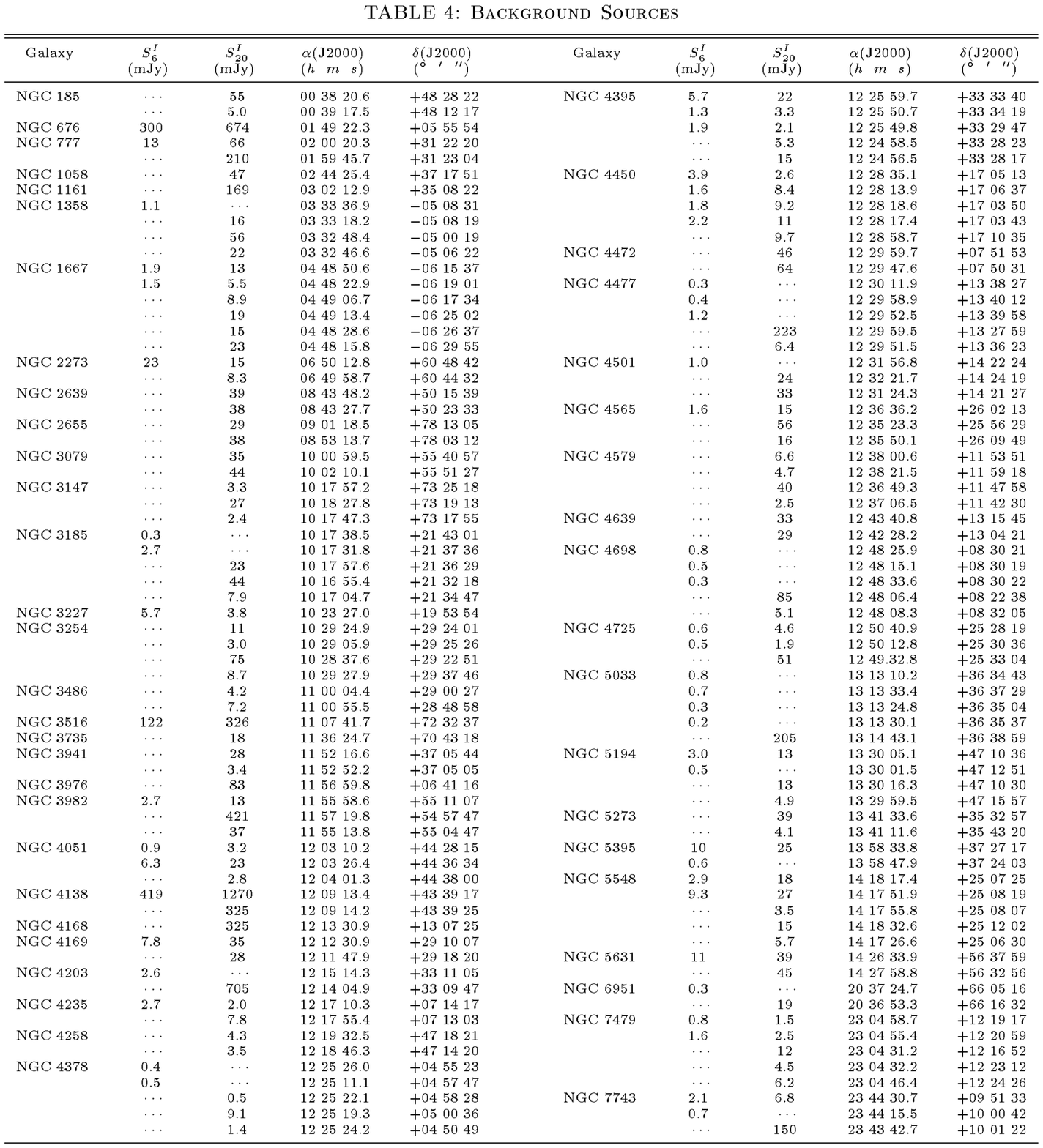,height=8.5in,width=7.0in,angle=0}

%Table 5 has bounding box = 50 120 590 650
%\clearpage
\psfig{file=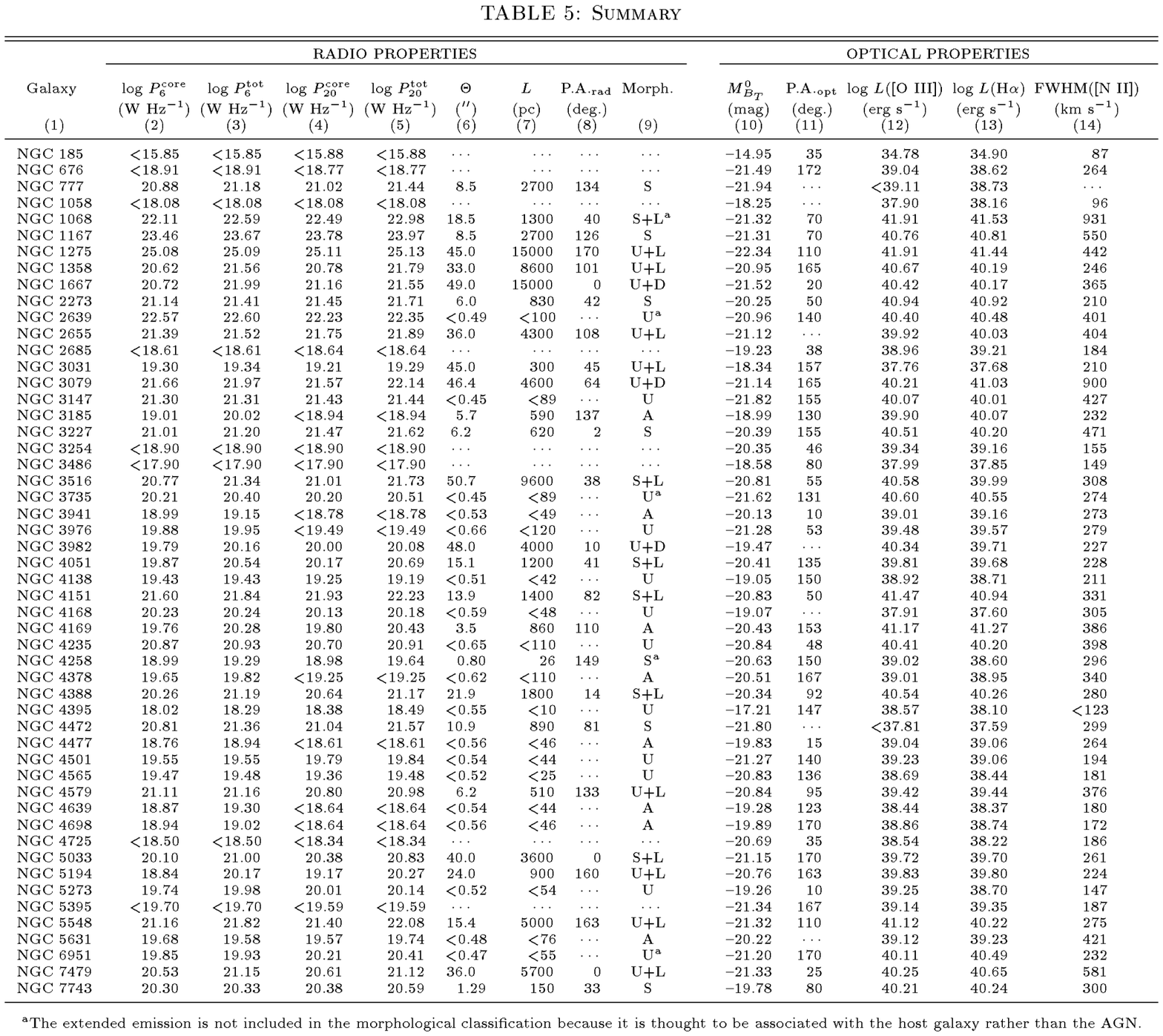,height=8.0in,width=7.0in,angle=0}

%FIGURES
\clearpage
\begin{figure}
\figurenum{1{\it a, b}}
\plotone{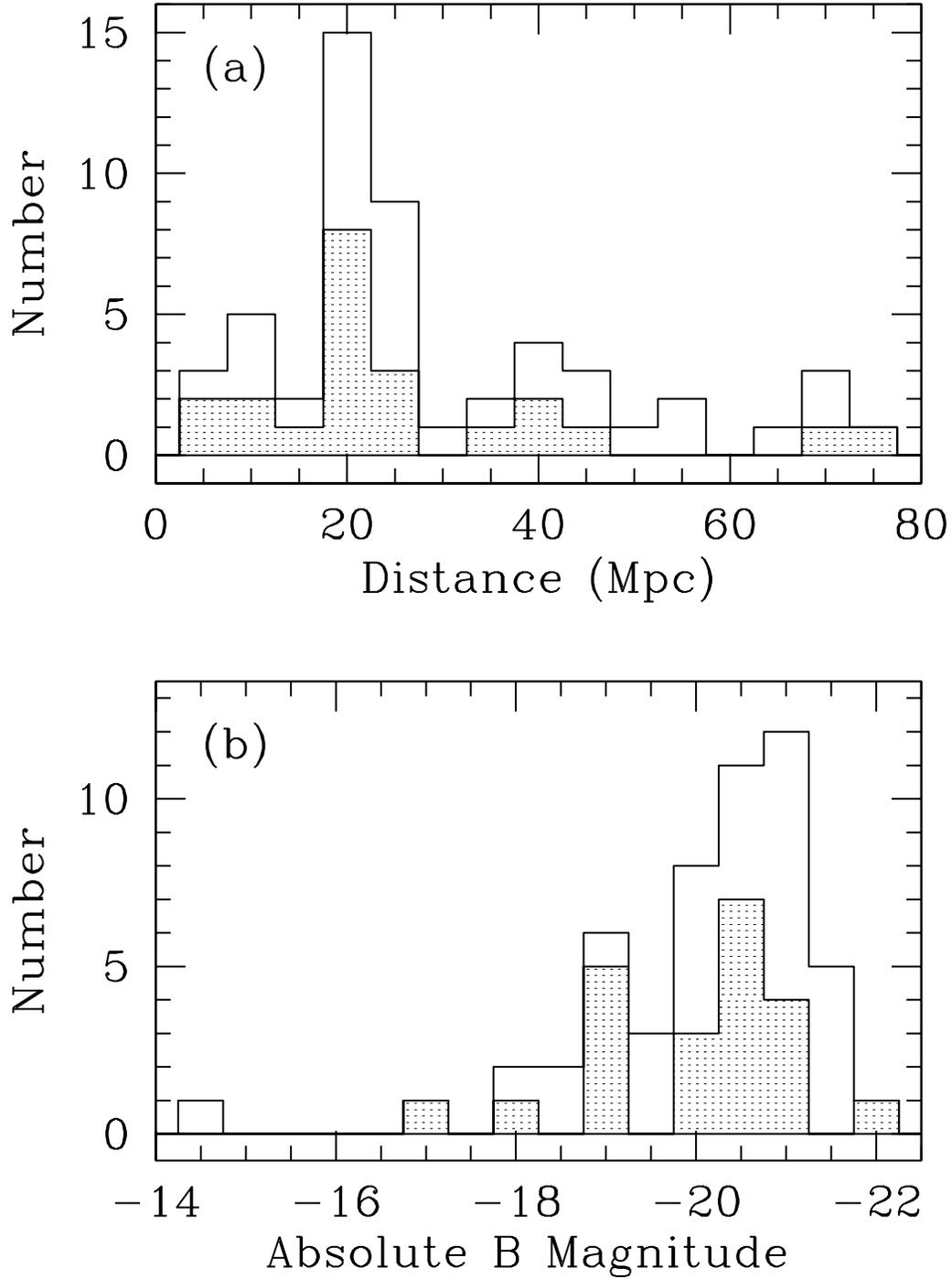}
\caption{
Number distributions of ({\it a\/}) distance and ({\it b\/}) total absolute 
blue magnitude (corrected for internal extinction) for the Palomar Seyfert 
sample.  The type~1 objects are shown as shaded histograms, and the total 
sample (types~1 and 2) is denoted by the unshaded histograms.  All the data 
are taken from Ho et al. (1997a) and are given in Table~1.
}
\end{figure}

\clearpage
\begin{figure}
\figurenum{1{\it c, d}}
\plotone{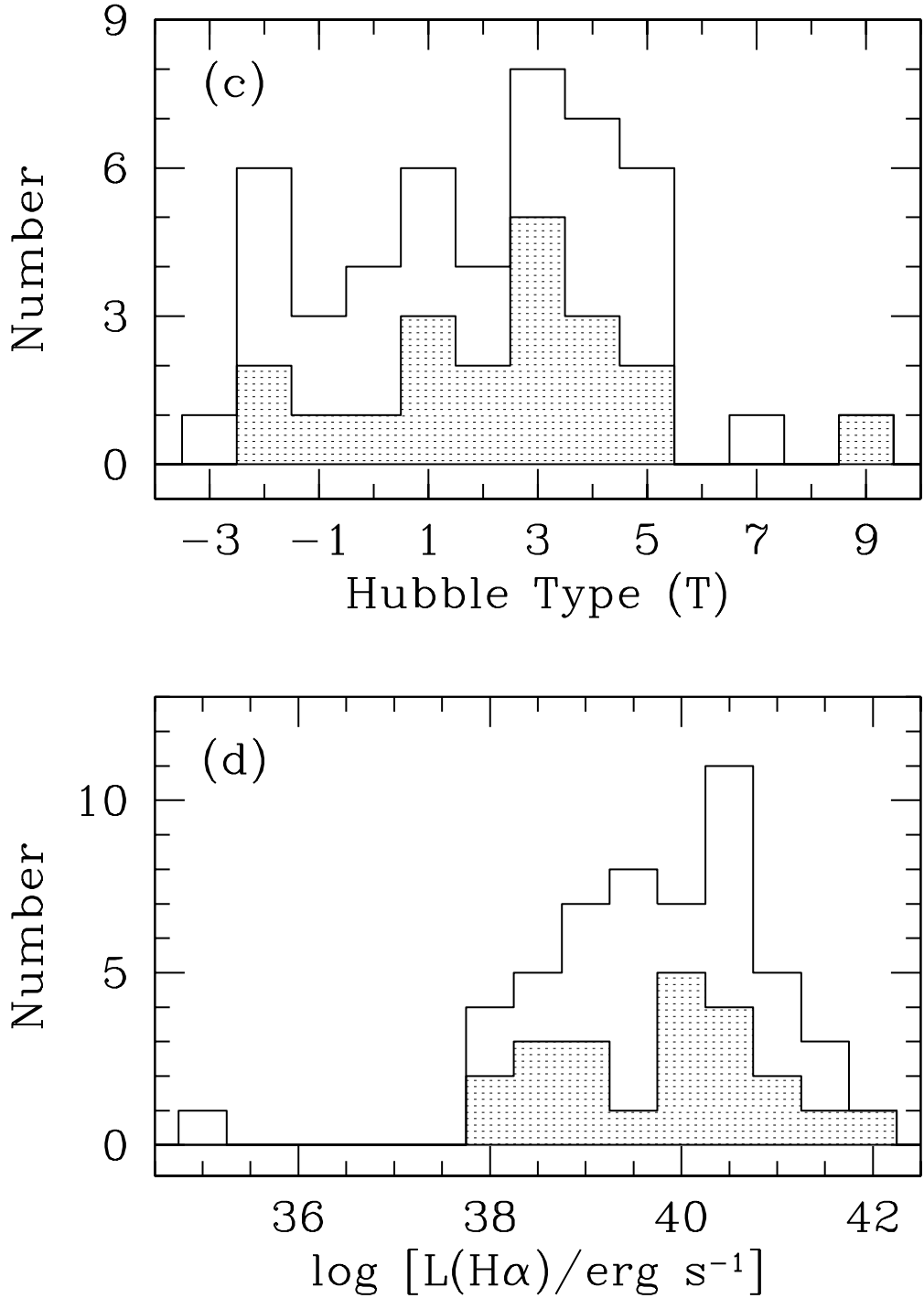}
\caption{
Number distributions of ({\it c\/}) morphological type index, $T$ 
(--3 to --1 = S0,  1 = Sa, 3 = Sb, 5 = Sc, 7 = Sd,  9 = Sm) and ({\it d\/}) 
extinction-corrected luminosity of the narrow H\al\ emission line for the 
Palomar Seyfert sample.  The type~1 objects are shown as shaded histograms, 
and the total sample (types~1 and 2) is denoted by the unshaded histograms.  
All the data are taken from Ho et al. (1997a) and are given in Table~1.
}
\end{figure}

\clearpage
\begin{figure}
\figurenum{2}
\centerline{{\Large{\it Figure shown separately as JPEG image.}}}
\vskip 1.0in
\caption{
Full-resolution 20~cm ({\it a\/}) and 6~cm ({\it b\/}) and tapered 6~cm 
({\it c\/}) maps of NGC~ 185, NGC~ 676, NGC~ 777, and NGC~1058.
}
\end{figure}

\clearpage
\begin{figure}
\figurenum{3}
\centerline{{\Large{\it Figure shown separately as JPEG image.}}}
\vskip 1.0in
\caption{
Full-resolution 20~cm ({\it a\/}) and 6~cm ({\it b\/}) and tapered 6~cm 
({\it c\/}) maps of NGC~1068, NGC~1167, NGC~1275, and NGC~1358.
}
\end{figure}

\clearpage
\begin{figure}
\figurenum{4}
\centerline{{\Large{\it Figure shown separately as JPEG image.}}}
\vskip 1.0in
\caption{
Full-resolution 20~cm ({\it a\/}) and 6~cm ({\it b\/}) and tapered 6~cm 
({\it c\/}) maps of NGC~1667, NGC~2273, NGC~2639, and NGC~2655.  Panel 
({\it c\/}) also shows the tapered 20~cm map of NGC~2655 (top).
}
\end{figure}

\clearpage
\begin{figure}
\figurenum{5}
\centerline{{\Large{\it Figure shown separately as JPEG image.}}}
\vskip 1.0in
\caption{
Full-resolution 20~cm ({\it a\/}) and 6~cm ({\it b\/}) and tapered 6~cm 
({\it c\/}) maps of NGC~2685, NGC~3031, NGC~3079, and NGC~3147.
}
\end{figure}

\clearpage
\begin{figure}
\figurenum{6}
\centerline{{\Large{\it Figure shown separately as JPEG image.}}}
\vskip 1.0in
\caption{
Full-resolution 20~cm ({\it a\/}) and 6~cm ({\it b\/}) and tapered 6~cm 
({\it c\/}) maps of NGC~3185, NGC~3227, NGC~3254, and NGC~3486.
}
\end{figure}

\clearpage
\begin{figure}
\figurenum{7}
\centerline{{\Large{\it Figure shown separately as JPEG image.}}}
\vskip 1.0in
\caption{
Full-resolution 20~cm ({\it a\/}) and 6~cm ({\it b\/}) and tapered 6~cm 
({\it c\/}) maps of NGC~3516, NGC~3735, NGC~3941, and NGC~3976.
}
\end{figure}

\clearpage
\begin{figure}
\figurenum{8}
\centerline{{\Large{\it Figure shown separately as JPEG image.}}}
\vskip 1.0in
\caption{
Full-resolution 20~cm ({\it a\/}) and 6~cm ({\it b\/}) and tapered 6~cm 
({\it c\/}) maps of NGC~3982, NGC~4051, NGC~4138, and NGC~4151.
}
\end{figure}

\clearpage
\begin{figure}
\figurenum{9}
\centerline{{\Large{\it Figure shown separately as JPEG image.}}}
\vskip 1.0in
\caption{
Full-resolution 20~cm ({\it a\/}) and 6~cm ({\it b\/}) and tapered 6~cm 
({\it c\/}) maps of NGC~4168, NGC~4169, NGC~4203, and NGC~4235.
}
\end{figure}

\clearpage
\begin{figure}
\figurenum{10}
\centerline{{\Large{\it Figure shown separately as JPEG image.}}}
\vskip 1.0in
\caption{
Full-resolution 20~cm ({\it a\/}) and 6~cm ({\it b\/}) and tapered 6~cm 
({\it c\/}) maps of NGC~4258, NGC~4378, NGC~4388, and NGC~4395.
}
\end{figure}

\clearpage
\begin{figure}
\figurenum{11}
\centerline{{\Large{\it Figure shown separately as JPEG image.}}}
\vskip 1.0in
\caption{
Full-resolution 20~cm ({\it a\/}) and 6~cm ({\it b\/}) and tapered 6~cm 
({\it c\/}) maps of NGC~4450, NGC~4472, NGC~4477, and NGC~4501.
}
\end{figure}

\clearpage
\begin{figure}
\figurenum{12}
\centerline{{\Large{\it Figure shown separately as JPEG image.}}}
\vskip 1.0in
\caption{
Full-resolution 20~cm ({\it a\/}) and 6~cm ({\it b\/}) and tapered 6~cm 
({\it c\/}) maps of NGC~4565, NGC~4579, NGC~4639, and NGC~4698.
}
\end{figure}

\clearpage
\begin{figure}
\figurenum{13}
\centerline{{\Large{\it Figure shown separately as JPEG image.}}}
\vskip 1.0in
\caption{
Full-resolution 20~cm ({\it a\/}) and 6~cm ({\it b\/}) and tapered 6~cm 
({\it c\/}) maps of NGC~4725, NGC~5033, NGC~5194, and NGC~5273.
}
\end{figure}

\clearpage
\begin{figure}
\figurenum{14}
\centerline{{\Large{\it Figure shown separately as JPEG image.}}}
\vskip 1.0in
\caption{
Full-resolution 20~cm ({\it a\/}) and 6~cm ({\it b\/}) and tapered 6~cm 
({\it c\/}) maps of NGC~5395, NGC~5548, NGC~5631, and NGC~6951.
}
\end{figure}

\clearpage
\begin{figure}
\figurenum{15}
\centerline{{\Large{\it Figure shown separately as JPEG image.}}}
\vskip 1.0in
\caption{
Full-resolution 20~cm ({\it a\/}) and 6~cm ({\it b\/}) and tapered 6~cm 
({\it c\/}) maps of NGC~7479 and NGC~7743.
}
\end{figure}

\clearpage
\begin{figure}
\figurenum{16{\it a, b}}
%\psfig{file=fig16-p1.ps,height=8.5in,angle=0}
%\plotone{fig16-p1.ps}
\centerline{{\Large{\it Figure shown separately as JPEG image.}}}
\vskip 1.0in
\caption{
Objects with detected linear polarization.   The 
contours show maps of total intensity, as in Fig. 2, and the
vectors represent linearly-polarized emission.
}
\end{figure}

\clearpage
\begin{figure}
\figurenum{16{\it c, d}}
%\psfig{file=fig16-p2.ps,height=8.5in,angle=0}
%\plotone{fig16-p2.ps}
\centerline{{\Large{\it Figure shown separately as JPEG image.}}}
\vskip 1.0in
\caption{
Objects with detected linear polarization.   The  
contours show maps of total intensity, as in Fig. 2, and the
vectors represent linearly-polarized emission. 
}
\end{figure}

\clearpage
\begin{figure}
\figurenum{16{\it e, f, g}}
%\psfig{file=fig16-p3.ps,height=8.5in,angle=0}
%\plotone{fig16-p3.ps}
\centerline{{\Large{\it Figure shown separately as JPEG image.}}}
\vskip 1.0in
\caption{
Objects with detected linear polarization.   The  
contours show maps of total intensity, as in Fig. 2, and the
vectors represent linearly-polarized emission. 
}
\end{figure}

\clearpage
\begin{figure}
\figurenum{16{\it h, i}}
%\psfig{file=fig16-p4.ps,height=8.5in,angle=0}
%\plotone{fig16-p4.ps}
\centerline{{\Large{\it Figure shown separately as JPEG image.}}}
\vskip 1.0in
\caption{
Objects with detected linear polarization.   The  
contours show maps of total intensity, as in Fig. 2, and the
vectors represent linearly-polarized emission. 
}
\end{figure}

\clearpage
\begin{figure}
\figurenum{16{\it j, k}}
%\psfig{file=fig16-p5.ps,height=8.5in,angle=0}
%\plotone{fig16-p5.ps}
\centerline{{\Large{\it Figure shown separately as JPEG image.}}}
\vskip 1.0in
\caption{
Objects with detected linear polarization.   The  
contours show maps of total intensity, as in Fig. 2, and the
vectors represent linearly-polarized emission. 
}
\end{figure}

\clearpage
\begin{figure}
\figurenum{16{\it l}}
%\psfig{file=fig16-p6.ps,height=8.5in,angle=0}
%\plotone{fig16-p6.ps}
\centerline{{\Large{\it Figure shown separately as JPEG image.}}}
\vskip 1.0in
\caption{
Objects with detected linear polarization.   The  
contours show maps of total intensity, as in Fig. 2, and the
vectors represent linearly-polarized emission. 
}
\end{figure}

\clearpage
\begin{figure}
\figurenum{17}
\plotone{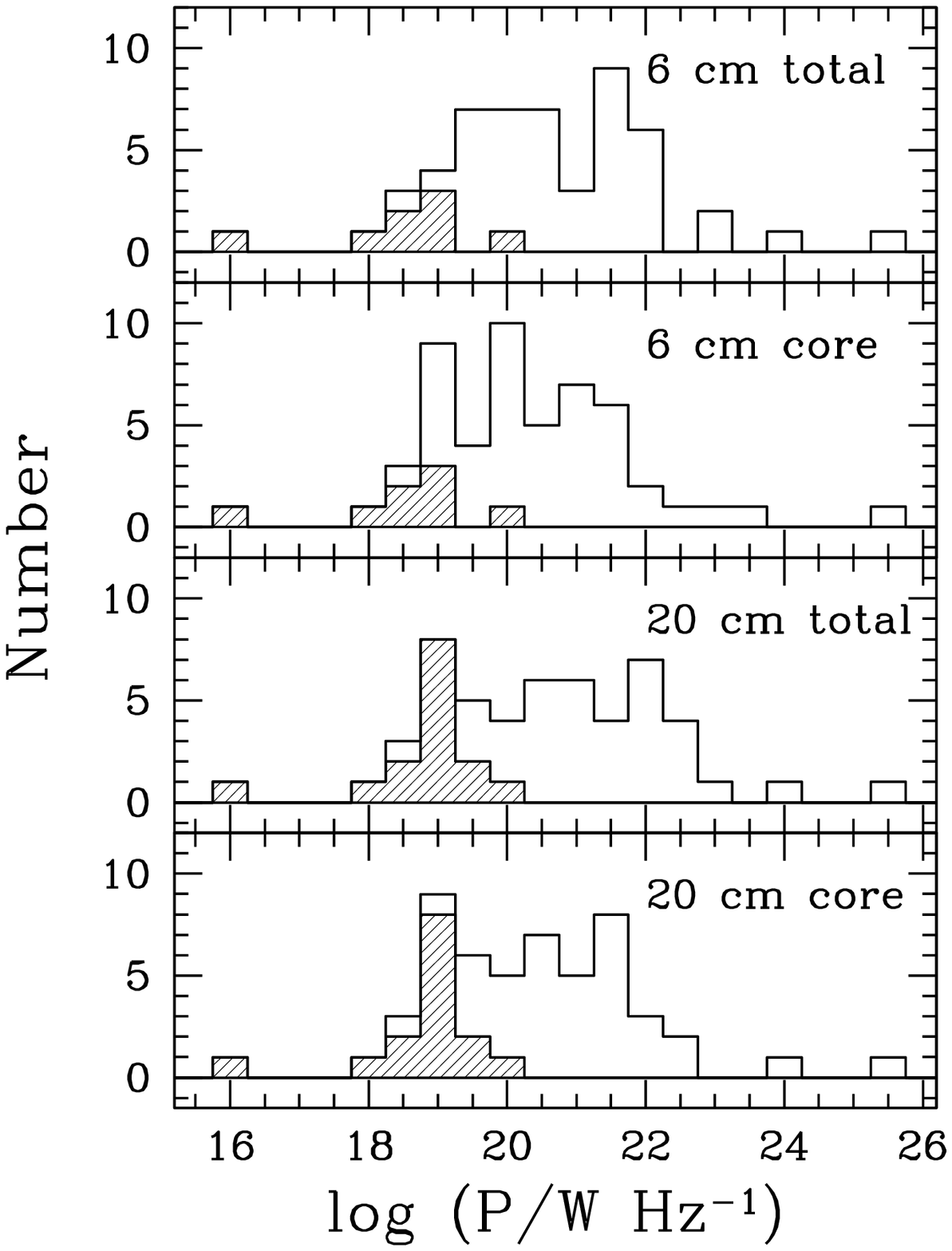}
\caption{
Number distributions of radio powers.  The panels show, from top to
bottom, measurements for the total integrated power at 6~cm, the peak core
power at 6~cm, the total integrated power at 20~cm, and the peak core
power at 20~cm.  The shaded regions of the histogram denote upper limits.
}
\end{figure}

\clearpage
\begin{figure}
\figurenum{18}
\plotone{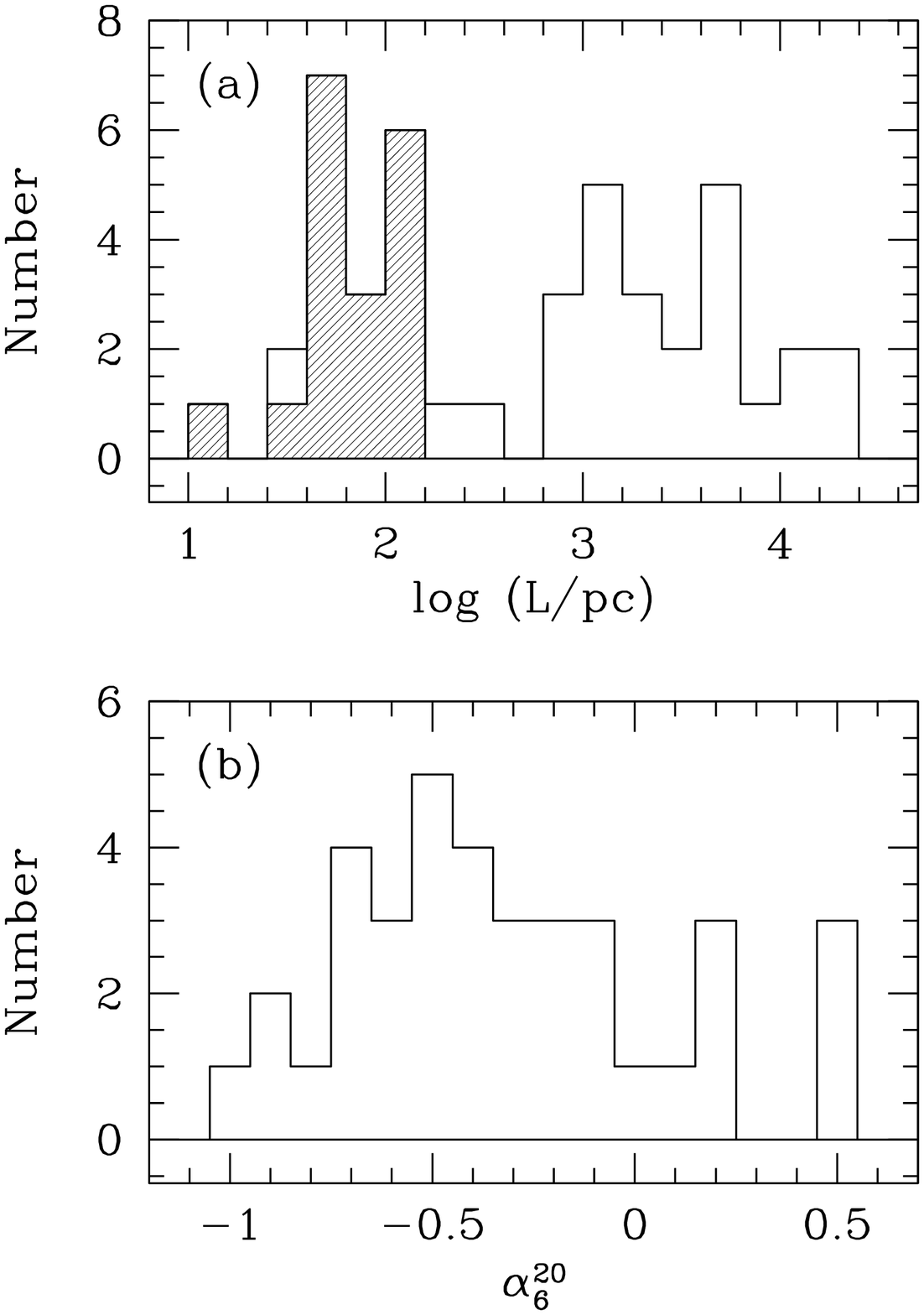}
\caption{
Number distributions of ({\it a\/}) linear source size and ({\it b\/}) 
spectral index between 6 and 20~cm.
The shaded regions of the histogram denote upper limits.
}
\end{figure}


\begin{thebibliography}{}

\bibitem{}
Argyle, R.~W., \& Eldridge, P. 1990, \mnras, 243, 504

\bibitem{}
Bartel, N., \etal 1982, \apj, 262, 556

\bibitem{} 
Barth, A.~J., Ho, L.~C., Filippenko, A.~V., \& Sargent, W.~L.~W. 1995, \aj,
110, 1009

\bibitem{} 
Barth, A.~J., Reichert, G.~A., Filippenko, A.~V., Ho, L.~C.,
Shields, J.~C., Mushotzky, R.~F., \& Puchnarewicz, E.~M. 1996, \aj, 112, 1829

\bibitem{} 
Baum, S.~A., O'Dea, C.~P., Dallacassa, D., de Bruyn, A.~G., \& Pedlar, A.
1993, \apj, 419, 553

\bibitem{} 
Becker, R.~H., White, R.~L., \& Helfand, D.~J. 1995, \apj, 450, 559

\bibitem{} 
Bicknell, G.~V., Dopita, M.~A., Tsvetanov, Z., \& Sutherland, R.~S. 1998,
\apj, 495, 680

\bibitem{} 
Bietenholz, M.~F., Bartel, N., \& Rupen, M.~P. 2000, \apj, 532, 895

\bibitem{} 
Birkinshaw, M., \& Davies, R.~L. 1985, \apj, 291, 32

\bibitem{} 
Booler, R.~V., Pedlar, A., \& Davies, R.~D. 1982, \mnras, 199, 229

\bibitem{} 
Bridle, A.~H., \& Fomalont, E.~B. 1978, \mnras, 185, 67P

\bibitem{} 
Bridle, A.~H., \& Schwab, F.~R. 1999, Synthesis Imaging in Radio Astronomy
II, ed. G.~B.  Taylor, C.~L. Carilli, \& R.~A. Perley (San Francisco: ASP),
371

\bibitem{} 
Browne, I.~W.~A., Patnaik, A.~R., Wilkinson, P.~N., \& Wrobel, J.~W. 1998,
\mnras, 293, 257

\bibitem{} 
Carral, P., Turner, J.~L., \& Ho, P.~T.~P. 1990, \apj, 362, 434

\bibitem{} 
Cecil, G. 1988, \apj, 329, 38

\bibitem{} 
Clark, B.~G. 1980, \aa, 89, 377

\bibitem{} 
Clements, E.~D. 1981, \mnras, 197, 829

\bibitem{} 
------. 1983, \mnras, 204, 811

\bibitem{} 
Cohen, R.~D. 1983, \apj, 273, 489

\bibitem{} 
Colina, L., Garc\'\i a Vargas, M.~L., Golz\'alez Delgado, R.~M., Mas-Hesse,
J.~M., P\'erez, E., Alberdi, A., \& Krabbe, A. 1997, \apj, 488, L71

\bibitem{} 
Collison, P.~M., Saikia, D.~J., Pedlar, A., Axon, D.~J., \& Unger, S.~W. 1994,
\mnras, 268, 203

\bibitem{} 
Condon, J.~J. 1980, \apj, 242, 894

\bibitem{} 
------. 1997, \pasp, 109, 166

\bibitem{} 
Condon, J.~J., \& Broderick, J.~J. 1988, \aj, 96, 30

\bibitem{} 
Condon, J.~J., Condon, M.~A., Gisler, G., \& Puschell, J.~J. 1982, \apj, 252,
102

\bibitem{} 
Condon, J.~J., Frayer, D.~T., \& Broderick, J.~J. 1991, \aj, 101, 362

\bibitem{} 
Cornwell, T.~J., \& Fomalont, E.~B. 1999, Synthesis Imaging in Radio
Astronomy II, ed. G.~B.  Taylor, C.~L. Carilli, \& R.~A. Perley (San
Francisco: ASP), 187

\bibitem{} 
Cotton, W.~D., Condon, J.~J., \& Arbizzani, E. 1999, \apjs, 125, 409

\bibitem{} 
Cowan, J.~J., Henry, R.~B.~C., \& Branch, D. 1988, \apj, 329, 116

\bibitem{}
Crane, P.~C., Giuffrida, B., \& Carlson, J.~B. 1976, \apj, 203, L113

\bibitem{}
Crane, P.~C., \& van der Hulst, J.~M. 1992, \aj, 103, 1146

\bibitem{} 
Dahari, O., \& De Robertis, M.~M. 1988, \apjs, 67, 249

\bibitem{} 
de~Bruyn, A.~G., Crane, P.~C., Price, R.~M., \& Carlson, J. 1976, \aa, 46, 243

\bibitem{} 
de Bruyn, A.~G., \& Wilson, A.~S. 1976, \aa, 53, 93

\bibitem{} 
------. 1978, \aa, 64, 433

\bibitem{} 
de Grijp, M.~H.~K., Keel, W.~C., Miley, G.~K., Goudfrooij, P., \& Lub, J.
1992, \aas, 96, 389
 
\bibitem{} 
de Grijp, M.~H.~K., Miley, G.~K., \& Lub, J. 1987, \aas, 70, 95
 
\bibitem{} 
de Grijp, M.~H.~K., Miley, G.~K., Lub, J., \& de Jong, T. 1985, \nat, 314, 240

\bibitem{} 
de Vaucouleurs, G., de Vaucouleurs, A., Corwin, H.~G., Jr., Buta, R.~J.,
Paturel, G., \& Fouqu\'e, R. 1991, Third Reference Catalogue of Bright
Galaxies (New York: Springer) 

\bibitem{} 
Dhawan, V., Kellermann, K.~I., \& Romney, J.~D. 1998, \apj, 498, L111

\bibitem{} 
Duric, N., \& Seaquist, E.~R. 1988, \apj, 326, 574

\bibitem{} 
Edelson, R. 1987, \apj, 313, 651

\bibitem{} 
Ekers, R.~D., \& Kotanyi, C.~G. 1978, \aa, 67, 47

\bibitem{} 
Eubanks, T.~M., et al. 1998, USNO - Ref Frame 1998-6, unpublished

\bibitem{} 
Fabbiano, G., Gioia, I.~M., \& Trinchieri, G. 1989, \apj, 347, 127

\bibitem{} 
Falcke, H., Wilson, A.~S., \& Simpson, C. 1998, \apj, 502, 199

\bibitem{} 
Fanaroff, B.~L., \& Riley, J.~M. 1974, \mnras, 167, 31P

\bibitem{} 
Fanti, C., Fanti, R., de Ruiter, H.~R., \& Parma, P. 1986, \aas, 65, 145

\bibitem{} 
Feigelson, E.~D., \& Nelson, P.~I. 1985, \apj, 293, 192

\bibitem{} 
Filippenko, A.~V., Ho, L.~C., \& Sargent, W.~L.~W. 1993, \apj, 410, L75

\bibitem{} 
Filippenko, A.~V., \& Sargent, W.~L.~W. 1989, \apj, 342, L11

%\bibitem{} 
%Fomalont, E.~B., Kellermann, K.~I., Wall, J.~V., \& Weistrop, D. 1984,
%Science, 225, 23

%\bibitem{} 
%Fomalont, E.~B., Windhorst, R.~A., Kristian, J.~A., \& Kellerman, K.~I.
%1991, \aj, 102, 1258

\bibitem{} 
Ford, H.~C., Crane, P.~C., Jacoby, G.~H., Lawrie, D.~G., \& van der Hulst,
J.~M. 1985, \apj, 293, 132

\bibitem{} 
Gallimore, J.~F., Baum, S.~A., \& O'Dea, C.~P. 1996, \apj, 458, 136

\bibitem{} 
Giuricin, G., Mardirossian, F., \& Mezzetti, M. 1988, \aa, 203, 39

\bibitem{} 
Giuricin, G., Mardirossian, F., Mezzetti, M., \& Bertotti, G. 1990, \apjs,
72, 551

\bibitem{} 
Gonz\'alez Delgado, R.~M., Heckman, T., Leitherer, C., Meurer, G., Krolik,
J.~H., Wilson, A.~S., Kinney, A.~L., \& Koratkar, A.~P. 1998, \apj, 505, 174

\bibitem{} 
Goodrich, R.~W. 1989, \apj, 342, 224

\bibitem{} 
Greenhill, L.~J., Herrnstein, J.~R., Moran, J.~M., Menten, K.~M., \&
Velusamy, T. 1997, \apj, 486, L15

\bibitem{} 
Heckman, T.~M., Balick, B., \& Crane, P.~C. 1980, \aas, 40, 295

\bibitem{} 
Heckman, T.~M., van Breugel, W.~J.~M., Miley, G.~K., \& Butcher, H.~R. 1983,
\aj, 88, 1077

\bibitem{} 
Herrnstein, J., Greenhill, L.~J., Moran, J.~M., Diamond, P., Inoue, M.,
Nakai, N., \& Miyoshi, M. 1998, \apj, 497, L69

\bibitem{} 
Herrnstein, J., Moran, J.~M., Greenhill, L.~J., Diamond, P., Miyoshi, M.,
Nakai, N., \& Inoue, M. 1997, \apj, 475, L17

\bibitem{} 
Ho, L.~C. 1996, in The Physics of LINERs in View of Recent Observations, 
ed. M. Eracleous et al. (San Francisco: ASP), 103

\bibitem{} 
Ho, L.~C., Filippenko, A.~V., \& Sargent, W.~L.~W. 1995, \apjs, 98, 477

\bibitem{} 
------. 1996, \apj, 462, 183

\bibitem{} 
------. 1997a, \apjs, 112, 315

\bibitem{} 
------. 1997b, \apj, 487, 568

\bibitem{} 
Ho, L.~C., Filippenko, A.~V., Sargent, W.~L.~W., \& Peng, C.~Y. 1997c, \apjs,
112, 391

\bibitem{} 
Ho, L.~C., Ptak, A., Terashima, Y., Kunieda, H., Serlemitsos, P.~J.,
Yaqoob, T., \& Koratkar, A.~P. 1999a, \apj, 525, 168

\bibitem{} 
Ho, L.~C., Van Dyk, S.~D., Pooley, G.~G., Sramek, R.~A., \& Weiler, K.~W.
1999b, \aj, 118, 843

\bibitem{} 
H\"ogbom, J.~A. 1974, \aas, 15, 417

\bibitem{} 
Huchra, J.~P., \& Burg, R. 1992, \apj, 393, 90

\bibitem{} 
Huchra, J.~P., Wyatt, W.~F., \& Davis, M. 1982, \aj, 87, 1628

\bibitem{} 
Hummel, E., Fanti, C., Parma, P., \& Schilizzi, R.~T. 1982, \aa, 114, 400

\bibitem{} 
Hummel, E., Pedlar, A., van der Hulst, J.~M., \& Davies, R.~D. 1985, \aas,
60, 293

\bibitem{} 
Hummel, E., \& Saikia, D.~J. 1991, \aa, 249, 43

\bibitem{} 
Hummel, E., van der Hulst, J.~M., \& Dickey, J.~M. 1984, \aa, 134, 207

\bibitem{} 
Hummel, E., van der Hulst, J.~M., Keel, W.~C., \& Kennicutt, R.~C., Jr. 1987,
\aas, 70, 517

\bibitem{}
Irwin, J.~A., English, J., \& Sorathia, B. 1999, \aj, 117, 2102

\bibitem{}
Kailey, W.~F., \& Lebofsky, M.~J. 1988, \apj, 326, 653

\bibitem{}
Kaufman, M., Bash, F.~N., Crane, P.~C., \& Jacoby, G.~H. 1996, \aj, 112, 1021

\bibitem{} 
Keel, W.~C. 1984, \apj, 282, 75

\bibitem{} 
Keel, W.~C., de Grijp, M.~H.~K., \& Miley, G.~K. 1988, \aa, 203, 250

\bibitem{} 
Keel, W.~C., de Grijp, M.~H.~K., Miley, G.~K., \& Zheng, W. 1994, \aa, 283, 791

\bibitem{} 
Keel, W.~C., \& Hummel, E. 1988, \aa, 194, 90

\bibitem{} 
Kinney, A.~L., Schmitt, H.~R., Clarke, C.~J., Pringle, J.~E., Ulvestad, J.~S., 
\& Antonucci, R.~R.~J. 2000, \apj, 537, 152

\bibitem{} 
Koski, A.~T. 1978, \apj, 223, 56

\bibitem{} 
Kukula, M.~J., Pedlar, A., Baum, S.~A., O'Dea, C.~P. 1995, \mnras, 276, 1262

\bibitem{} 
Laurent-Muehleisen, S.~A., Kollgaard, R.~I., Ryan, P.~J., Feigelson, E.~D.,
Brinkmann, W., \& Siebert, J. 1997, \aas, 122, 235

\bibitem{} 
Lira, P., Lawrence, A., O'Brien, P., Johnson, R.~A., Terlevich, R., \&
Bannister, N. 1999, \mnras, 304, 109

\bibitem{} 
Low, F.~J., Huchra, J.~P., Kleinmann, S.~G., \& Cutri, R.~M. 1988, \apj, 327, L41

\bibitem{} 
Ma, C., Arias, E.~F., Eubanks, T.~M., Fey, A.~L., Gontier, A.-M., Jacobs, 
C.~S., Sovers, O.~J., Archinal, B.~A., \& Charlot, P. 1998, \aj, 116, 516

\bibitem{} 
Maiolino, R., \& Rieke, G.~H. 1995, \apj, 454, 95

\bibitem{} 
Meurs, E.~J.~A., \& Wilson, A.~S. 1981, \aas, 45, 99

\bibitem{} 
------. 1984, \aa, 136, 206

\bibitem{} 
Miyaji, T., Wilson, A.~S., \& P\'erez-Fournon, I. 1992, \apj, 385, 137

\bibitem{} 
Moran, E.~C., Filippenko, A.~V., Ho, L.~C., Shields, J.~C., Belloni, T.,
Comastri, A., Snowden, S.~L., \& Sramek, R.~A. 1999, \pasp, 111, 801

\bibitem{} 
Morganti, R., Tsvetanov, Z.~I., Gallimore, J., \& Allen, M.~G. 1999, \aas,
137, 457

\bibitem{} 
Mundell, C.~G., Holloway, A.~J., Pedlar, A., Meaburn, J., Kukula, M.~J., \&
Axon, D.~J. 1995, \mnras, 275, 67

\bibitem{} 
Mundell, C.~G., Wilson, A.~S., Ulvestad, J.~S., \& Roy, A.~L. 2000, \apj, 529,
816

\bibitem{} 
Nagar, N.~M., Wilson, A.~S., Mulchaey, J.~S., \& Gallimore, J.~F. 1999, \apjs,
120, 209

\bibitem{} 
Norris, R.~P., Allen, D.~A., Sramek, R.~A., Kesteven, M.~J., \& Troup, E.~R.
1990, \apj, 359, 291

\bibitem{} 
Norris, R.~P., Kesteven, M.~J., Allen, D.~A., \& Troup, E.~R. 1988, \mnras,
234, 51P

\bibitem{} 
Osterbrock, D.~E. 1981, \apj, 249, 462

\bibitem{} 
Osterbrock, D.~E., \& De Robertis, M.~M. 1985, \pasp, 97, 1129

\bibitem{} 
Osterbrock, D.~E., \& Pogge, R.~W. 1985, \apj, 297, 166

\bibitem{} 
Patnaik, A.~R., Browne, I.~W.~A., Wilkinson, P.~N., \& Wrobel, J.~M. 1992,
\mnras, 254, 665

\bibitem{} 
Pedlar, A., Ghataure, H.~S., Davies, R.~D., Harrison, B.~A., Perley, R.,
Crane, P.~C., \& Unger, S.~W. 1990, \mnras, 246, 477

\bibitem{} 
Pedlar, A., Kukula, M.~J., Longley, D.~P.~T., Muxlow, T.~W.~B., Axon, D.~J., 
Baum, S., O'Dea, C., \& Unger, S.~W. 1993, \mnras, 263, 471

\bibitem{} 
Phillips, M.~M., Charles, P.~A., \& Baldwin, J.~A. 1983, \apj, 266, 485

\bibitem{} 
Piccinotti, G., Mushotzky, R.~F., Boldt, E.~A., Holt, S.~S., Marshall, 
F.~E., Serlemitsos, P.~J., \& Shafer, R.~A. 1982, \apj, 253, 485

\bibitem{} 
Roy, A.~L., Norris, R.~P., Kesteven, M.~J., Troup, E.~R., \& Reynolds, J.~E.
1994, \apj, 432, 496

\bibitem{} 
Rush, B., Malkan, M.~A., \& Edelson, R.~A. 1996, \apj, 473, 130

\bibitem{} 
Rush, B., Malkan, M.~A., \& Spinoglio, L. 1993, \apjs, 89, 1

\bibitem{} 
Sadler, E.~M., Jenkins, C.~R., \& Kotanyi, C.~G. 1989, \mnras, 240, 591

\bibitem{} 
Sadler, E.~M., Slee, O.~B., Reynolds, J.~E., \& Roy, A.~L. 1995, \mnras, 276,
1373

\bibitem{} 
Saikia, D.~J., Pedlar, A., Unger, S.~W., \& Axon, D.~J. 1994, \mnras, 270, 46

\bibitem{} 
Sandage, A.~R., \& Tammann, G.~A. 1981, A Revised Shapley-Ames Catalog of
Bright Galaxies (Washington, DC: Carnegie Inst. of Washington)

\bibitem{} 
Sandage, A.~R., Tammann, G.~A., \& Yahil, A. 1979, \apj, 232, 352

\bibitem{} 
Schmitt, H.~R., Ulvestad, J.~S., Antonucci, R.~R.~J., \& Kinney, A.~L.
2001, \apj, in press

\bibitem{} 
Silver, C.~S., Taylor, G.~B., \& Vermeulen, R.~C. 1998, \apj, 502, 229

\bibitem{} 
Slee, O.~B., Sadler, E.~M., Reynolds, J.~E., \& Ekers, R.~D. 1994, \mnras, 269,
928

\bibitem{} 
Spinoglio, L., \& Malkan, M.~A. 1989, \apj, 342, 83

\bibitem{} 
Sramek, R. 1992, in Relationships between AGNs and Starburst Galaxies, ed. 
A.~V. Filippenko (San Francisco: ASP), 273

\bibitem{} 
Stone, J.~L., Jr., Wilson, A.~S., \& Ward, M.~J. 1988, \apj, 330, 105

\bibitem{} 
Thean, A., Pedlar, A., Kukula, M.~J., Baum, S.~A., \& O'Dea, C.~P. 2000, 
\mnras, 314, 573

\bibitem{} 
Thompson, A.~R., Clark, B.~G., Wade, C.~M., \& Napier, P.~J. 1980, \apjs, 
44, 151

\bibitem{} 
Tonry, J., \& Davis, M. 1979, \aj, 84, 1511

\bibitem{} 
Tully, R.~B. 1988, Nearby Galaxies Catalog (Cambridge: Cambridge Univ. Press)

\bibitem{} 
Turner, J.~L., \& Ho, P.~T.~P. 1994, \apj, 421, 122

\bibitem{} 
Turner, K.~C., Helou, G., \& Terzian, Y. 1988, \pasp, 100, 452

\bibitem{} 
Ulvestad, J.~S. 1986, \apj, 310, 136

\bibitem{} 
Ulvestad, J.~S., Antonucci, R.~R.~J., \& Goodrich, R.~W. 1995, \aj, 109, 81:

\bibitem{} 
Ulvestad, J.~S., \& Ho, L.~C. 2001, in preparation

\bibitem{} 
Ulvestad, J.~S., Neff, S.~G., \& Wilson, A.~S. 1987, \aj, 93, 22

\bibitem{} 
Ulvestad, J.~S., Roy, A.~L., Colbert, E.~J.~M., \& Wilson, A.~S. 1998, \apj,
496, 196

\bibitem{} 
Ulvestad, J.~S., \& Wilson, A.~S. 1984a, \apj, 278, 544

\bibitem{} 
------. 1984b, \apj, 285, 439

\bibitem{} 
------. 1989, \apj, 343, 659

\bibitem{} 
Ulvestad, J.~S., Wilson, A.~S., \& Sramek, R.~A. 1981, \apj, 247, 419

\bibitem{} 
Unger, S.~W., Lawrence, A., Wilson, A.~S., Elvis, M., \& Wright, A.~E. 1987,
\mnras, 228, 521

\bibitem{} 
van Albada, T.~S., \& van der Hulst, J.~M. 1982, \aa, 115, 263

\bibitem{} 
van der Hulst, J.~M., Crane P.~C., \&  Keel, W.~C. 1981, \aj, 86, 1175
(Erratum:  1983, \aj, 88, 138)

\bibitem{} 
van Moorsel, G., Kemball, A., \& Greisen, E. 1996, in Astronomical Data  
Analysis Software and Systems V, ed. G.~H. Jacoby \& J. Barnes (San Francisco:
ASP), 37

\bibitem{} 
Veilleux, S., Kim, D.-C., \& Sanders, D.~B. 1999, \apj, 522, 113

\bibitem{} 
V\'eron, P. 1979, \aa, 78, 46

\bibitem{} 
Vila, M.~B., Pedlar, A., Davies, R.~D., Hummel, E., \& Axon, D.~J. 1990,
\mnras, 242, 379

\bibitem{} 
Walker, R.~C., Dhawan, V., Romney, J.~D., Kellerman, K.~I., \& Vermeulen,
R.~C. 2000, \apj, 530, 233

\bibitem{} 
Wardle, J.~F.~C., \& Kronberg, P.~P. 1974, \apj, 194, 249

\bibitem{} 
Weiler, K.~W., Sramek, R.~A., Panagia, N., van der Hulst, J.~M., \&
Salvati, M. 1986, \apj, 301, 790

\bibitem{} 
Whittle, M. 1985, \mnras, 213, 33

\bibitem{} 
------. 1992, \apjs, 79, 49

\bibitem{} 
Wilkinson, P.~N., Browne, I.~W.~A., Patnaik, A.~R., Wrobel, J.~M., \&
Sorathia, B. 1998, \mnras, 300, 790

\bibitem{} 
Wilson, A.~S. 1997, in IAU Colloq. 159, Emission Lines in Active Galaxies:
New Methods and Techniques, ed. B.~M.  Peterson, F.-Z. Cheng, \& A.~S. Wilson
(San Francisco: ASP), 264

\bibitem{} 
Wilson, A.~S., \etal 1998, \apj, 505, 587

\bibitem{} 
Wilson, A.~S., \& Meurs, E.~J.~A. 1982, \aas, 50, 217

\bibitem{} 
Wilson, A.~S., \& Ulvestad, J.~S. 1982a, \apj, 260, 56

\bibitem{} 
------. 1982b, \apj, 263, 576

\bibitem{} 
------. 1983, \apj, 275, 8

\bibitem{} 
Wilson, A.~S., \&  Willis, A.~J. 1980, \apj, 240, 429

\bibitem{} 
Windhorst, R.~A., Miley, G.~K., Owen, F.~N., Kron, R.~G., \& Koo, D.~C. 1985,
\apj, 289, 494

\bibitem{} 
Wrobel, J.~M. 2000, \apj, 531, 716

\bibitem{} 
Wrobel, J.~M., \& Heeschen, D.~S. 1991, \aj, 101, 148

\end{thebibliography}
\end{document}